\documentclass[trackchanges, preprint2, twocolappendix, resetfootnote]{aastex7}

\definecolor{royalblue}{HTML}{4169e1}
\definecolor{forestgreen}{HTML}{228b22}
\definecolor{darkorange}{HTML}{ff8c00}
\definecolor{darkorchid}{HTML}{9932cc}

\begin{document}

\title{Old Disks Die Hard: How Does AGN Feedback Suppress Disk Formation in Milky Way Mass Galaxies?} 

\author[0000-0002-1218-659X]{Patricia Fofie}
\affiliation{University of California Irvine, Physics \& Astronomy Department}
\email[show]{pfofie@uci.edu}  

\author[0000-0003-4298-5082]{James S. Bullock}
\affiliation{University of Southern California, Physics \& Astronomy Department}
\email{jamesbul@usc.edu}

\author[0000-0002-3977-2724]{Sarah Wellons}
\affiliation{Wesleyan University, Department of Astronomy}
\email{swellons@wesleyan.edu}

\author[0000-0002-3430-3232]{Jorge Moreno}
\affiliation{Pomona College, Physics \& Astronomy Department}
\email{jorge.moreno@pomona.edu}

\author[0000-0002-5908-737X]{Francisco J. Mercado}
\affiliation{Pomona College, Physics \& Astronomy Department}
\email{francisco.mercado@usc.edu}

\author[0000-0003-0603-8942]{Andrew Wetzel}
\affiliation{University of California Davis, Department of Physics \& Astronomy}
\email{awetzel@ucdavis.edu}

\begin{abstract}
The connection between a galaxy's visual shape and its star formation rate is one of the oldest established trends in galaxy evolution, yet the origin of this trend remains unsettled. We analyze two Milky-Way mass halos simulated with FIRE-2 galaxy formation physics, each run without AGN and with up to three implementations of AGN feedback.  We use them to study how and why AGN star formation suppression affects morphological evolution over cosmic time. Both runs without AGN feedback produce prominent, thin, star-forming spiral disks at $z=0$.  Each of the AGN runs has less late-time star formation, and ends up with a higher spheroid fraction and a lower thin-disk mass fraction.  In two instances, the AGN runs produce quenched lenticular galaxies with no cool gas at $z=0$. The main reason for these morphological differences is that AGN feedback becomes effective {\em just after} the onset of disk formation, at ``spin up," in every run. This time-differentiated impact suppresses the formation of disk stars, especially thin-disk stars, preferentially because thin-disk formation occurs late. The spheroidal components are assembled early in all runs, before AGN feedback is effective, and are therefore similar in mass and size across runs. 
\end{abstract}


\section{Introduction} \label{sec:intro}
The relationship between morphology and star formation in galaxy populations is among the core phenomenological correlations in galaxy formation \citep{Holmberg1958, Roberts1963, Kennicutt1983}. In the local universe, massive galaxies ($M_{\star} \gtrsim 10^{11} \,{\rm M}_\odot$) are typically quenched with lenticular or elliptical morphologies while lower-mass field galaxies ($M_{\star} \lesssim 10^{10}~{\rm M}_\odot$) are usually star-forming with spiral, irregular, or spheroidal morphologies \citep{Roberts94, Fioc1999, Baldry2012, Geha2012, Kelvin2014, Mahajan2015, Bluck2022}. At intermediate masses ($M_{\star} \sim 10^{10.5} \,\rm M_{\odot}$), galaxies occupy an overlapping population \citep[][]{Strateva2001, Kauffmann2003, Driver2006, FraserMcKelvieCortese2022, Bluck2022}: many lie on the quenched/red sequence and are early-type (often called \small{`red \& dead'}), while many others remain on the star-forming/blue cloud and are disk-dominated, which we dub \small{`blue \& new'} galaxies. 

While these correlations are strong, with the vast majority of quiescent galaxies having prominent bulges and bulge mass correlated with central black hole mass \citep[see][]{Bluck2014}, there are mixed cases that could provide clues to broader physical origins.  Specifically, some quenched galaxies exhibit significant disks, while others are bulge-dominated and show ongoing star formation, which could represent galaxies transitioning from disky \small{`blue \& new'} to spheroidal \small{`red \& dead'} galaxies \citep{Bell2012, Koyama2025}. 
 
It is widely believed that massive galaxies are quenched via supermassive black hole (SMBH) feedback \citep[e.g.][]{Silk&Rees1998, Schawinski2009, 2017Bower, 2017Harrison, Bluck2022, French2023}, and that the feedback itself is important for driving the observed correlations between black hole mass and galaxy properties like spheroid mass and/or velocity dispersion \citep[e.g.,][]{Magorrian1998, FerrareseMerritt2000, Gebhardt2000, Fabian2012, Kormendy2013}. Supermassive black holes at the centers of massive galaxies accrete material from the host, and the resulting emission/radiation, observed as active galactic nuclei (AGN), can interact with the gas in the interstellar medium (ISM) and circumgalactic medium (CGM) of the host galaxy \citep[e.g.,][]{DiMatteo05, Hopkins2008}. When coupled, AGN feedback can disrupt the host galaxy gas reservoir by imparting energy and momentum via outflows and heating. Outflows can expel the cold gas that fuels star formation, while the heating prevents the remaining gas from cooling efficiently to form new stars. Various simulations and models require some form of energy and momentum injection from AGN feedback to regulate both the star formation activity in the host galaxy and the growth of the central SMBH, and to reproduce observed galactic scaling relations, \citep[see][for a review]{2015Somerville&Dave}. 

Though there is near consensus that AGN feedback is needed for galaxy quenching at the massive end ($M_{\star} > 10^{10.5} \,\rm M_{\odot}$), the precise way in which black hole quenching shapes galaxy morphology remains a topic of debate.  An important clue comes from studies at $z>2$, which suggest that significant bulge growth precedes quenching in massive galaxies \citep{Lang2014, Borgohain2025}. This also suggests that disks are quenched preferentially.

One clear connection between the absence of disks and AGN activity occurs in the limited case where major mergers trigger rapid feeding of SMBHs. In this circumstance, quenching and the destruction of disks arise naturally: mergers scramble stellar orbits and also trigger AGN to expel gas and quench star formation \citep{Hopkins2008}. However, AGN feedback is not necessarily limited to mergers and not all mergers trigger AGN \citep[see e.g.,][]{Cisternas2011, Treister2012}. This suggests that other mechanisms are needed to provide a full picture.

Of course, even in the absence of discrete mergers, high-accretion rates can trigger ``quasar'' modes that launch fast winds and expel disk ISM gas directly \citep{SilkRees1998, DiMatteo2005}.  Coherent gas disks are therefore destroyed {\em in situ}, and without fuel, star formation quenches as well. However, since accretion rates are expected to be highest at early times and observed quasar activity is not common at low redshift, preventing the regrowth of new disks remains a problem to solve in this scenario.

AGN can also deliver preventative feedback by heating halo gas and ``starving" galaxies slowly \citep{Bower2006}.  Specifically, low-accretion-rate SMBHs can keep the CGM hot, prevent cooling, and also block the inflow of cold streams \citep{Fabian2012}. By halting the accretion of incoming cold gas, AGN can naturally halt disk growth \citep{Croton2006}.
In support of this preventative feedback idea is the substantial population of observed red, passive disks \citep{Bundy2010, Toft2017}.

All of these scenarios are likely at play to some degree.  Specifically, fast (merger-driven or rapid accretion-driven) quenching and slow (halo-heated) quenching seem to be required to reproduce the variety in galaxy populations \citep[see, e.g.][]{Appleby2020}. Further, mergers, independent of AGN quenching, may be needed to explain differences in black hole to spheroid scaling relations among lenticular bulge and elliptical populations \citep[][]{GS23}.
Further, among the most massive disks where \textit{in-situ} star formation is shut down by AGN feedback, there may be cases when subsequent galaxy mass growth occurs predominantly via \textit{ex-situ} accretion through dry mergers. Without cold gas to damp orbital energy and form new stars on circular orbits, stellar mergers randomize existing stellar orbits into spheroidal early-type galaxies.

This study focuses on simulations of Milky Way (MW) mass galaxies, which are below the mass scale where dry mergers should be common and also near the mass scale of morphological bimodality \citep{Taylor2015}.  At this scale, the local galaxy population includes both quenched, late-type systems and disk-dominated, star-forming galaxies, with green-valley galaxies occupying the transitional regime between these populations \citep{Kelvin2014, Smith2022}. This makes MW-mass galaxies an especially interesting regime for studying the relationship between quenching and disk formation. For example, \citet{Davis2018} find that early-type galaxies at the MW scale have more massive SMBHs than late-type galaxies of the same stellar mass, suggesting that black hole growth is linked to those morphological differences in a direct way. 

Previous studies exploring disk formation using Feedback In Realistic Environments (FIRE) \footnote{https://fire.northwestern.edu/} simulations of MW-like galaxies {\em without} AGN feedback show that these galaxies build up their morphological structure over three phases: 1) an early period where star formation is bursty and stars form on fairly radial, spheroidal type orbits; 2) a ``spin-up" phase during which star formation remains bursty but the ISM begins to resemble a turbulent, thick disk; and 3) a late time-steady phase of star formation with thin-disk kinematics \citep[][]{Stern2021, Yu2021, Yu2023, Gurvich2023, McCluskey2025, Myrtaj2026, Bellardini2026}. Similar trends are apparent for galaxies formed in the TNG50 simulations \citep{semenov2024formation, Semenov2026DiskFormation}.   This picture is supported by studies of the Milky Way using Gaia XP and APOGEE data \citep{Belokurov2022, Chandra2024}, showing that the same three phases are present when stellar angular momentum or orbital circularity is studied as a function of stellar metallicity, used as an approximate proxy for stellar age. Natural questions to ask include whether AGN feedback affects these phases and if these phases can help us understand why AGN affects morphological structure.

Of particular relevance for this work is the idea that the third, thin-disk phase is enabled by the late-time virialization of the inner CGM, which allows the disk to be fed by gas with very well-aligned angular momentum from a calm and quasi-stable rotating cooling flow \citep{Stern2021, Hafen2022, Yu2023, Stern2024, Myrtaj2026}. This suggests that if AGN feedback were to disrupt the stable CGM at late times, it may preferentially impede thin-disk formation.

Recently, \cite{Wellons2023} analyzed a suite of FIRE-2 simulations with varying AGN feedback models and implementations, finding models that produce quenched elliptical galaxies at high mass and that reproduce observed relationships between SMBH mass and galaxy properties across halo mass ($M_{\rm vir}  = 10^{11} - 10^{13} \,\rm M_{\odot}$), including $M_{\rm BH} - \sigma$ and the Stellar Mass to Halo Mass relation (SMHM). An analysis of how and why morphology responds to star formation suppression from AGN feedback has not been performed in detail using FIRE-2 simulations. 

To elucidate the link between star formation and morphology under AGN feedback, specifically the link between the preferential suppression of disk formation and star formation generally, we analyze a subset of Milky-Way mass FIRE-2 galaxies with and without AGN feedback, which both suppresses star formation and affects end-state morphology. Using a kinematic proxy to characterize morphology, we examine how AGN feedback influences morphology and the impact of different implementations of SMBH feedback on the overall shape and structure of the simulated galaxies. 

We summarize the general physics of the simulation suite and the AGN feedback implementations in Section \ref{sec:data}. In Section \ref{sec:results}, we present the results of our analysis, including mock images of the galaxies at the present day and an analysis of disk formation over cosmic time. Lastly, in Section \ref{sec:discussion}, we discuss the additional impact of the specific AGN implementation on morphology.

\begin{deluxetable*}{cccccccccc}
\tabletypesize{\footnotesize}
\tablecaption{Summary table of the relevant properties of the seven simulations analyzed in this paper at $z=0$. AGN implementation lists how BH feedback is injected into the simulation. Particle spawning produces high-resolution particles containing the feedback energy and momentum, while the gas injection method injects the feedback energy directly into surrounding gas cells; see Section \ref{subsec:sims} for details. Columns 4-5 provide the total stellar mass and total gas mass within 20 kpc at $z=0$. Column 6 lists the final central BH mass.  Morphological indicators are listed in columns 7-9: the thin disk stellar mass fraction, the thick disk stellar mass fraction, and the spheroidal stellar mass fraction. The final column provides ``spin-up" lookback time, which signifies the transition from disordered kinematics to a sustained rise towards coherent spin; see Section \ref{subsec:circparam} for details.\label{tab:one}} 
\tablecolumns{9} 
\tablehead{
  \multicolumn{3}{c}{\textbf{AGN Implementation}} &
  \multicolumn{3}{c}{\textbf{Baryonic Mass}} &
  \multicolumn{3}{c}{\textbf{Morphology}} &
   \colhead{\textbf{Spin Up}} \\
  \colhead{Name} &
  \colhead{Feedback} &
  \colhead{Geometry} &
  \colhead{$M_{\rm \star}$} &
  \colhead{$M_{\rm gas}$} &
  \colhead{$M_{\rm BH}$} &
  \colhead{$f_{\rm thin}$} &
  \colhead{$f_{\rm thick}$} &
  \colhead{$f_{\rm sph}$} &
  \colhead{$t_{\rm lbt}$} \\[-5pt]
  \colhead{} &
  \colhead{Method} &
  \colhead{} &
  \colhead{$10^{10}$ $\rm M_{\odot}$} &
  \colhead{$10^{10}$ $\rm M_{\odot}$} &
  \colhead{$10^{6}$ $\rm M_{\odot}$} &
  \colhead{} &
  \colhead{} &
  \colhead{} &
  \colhead{Gyr} \\[-13pt]
}
\startdata
\cutinhead{\textbf{m12f ($M_{\rm halo} = 1.4\times10^{12}$ $\rm M_{\odot}$}) }
{\color[HTML]{4169E1} No}    & n/a            & n/a        & $8.4$ & $2.0$ & n/a   & 0.40 & 0.38 & 0.22 & 10.1 \\
{\color[HTML]{228B22} Push}  & gas injection  & isotropic  & $3.3$ & $2.0$ & $9.5$ & 0.24 & 0.45 & 0.31 & 10.9 \\
{\color[HTML]{FF8C00} Jet}   & particle spawn & collimated & $2.3$ & $0.8$ & $6.1$ & 0.16 & 0.42 & 0.43 & 11.0 \\
\cutinhead{\textbf{m12i ($M_{\rm halo} = 1.0\times10^{12}$ $\rm M_{\odot}$}) }
{\color[HTML]{4169E1} No}    & n/a            & n/a        & $5.9$ & $1.4$  & n/a   & 0.35 & 0.41 & 0.24 & 7.8 \\
{\color[HTML]{228B22} Push}  & gas injection  & isotropic  & $3.0$ & $0.4$  & $6.4$ & 0.26 & 0.44 & 0.30 & 8.6 \\
{\color[HTML]{FF8C00} Jet}   & particle spawn & collimated & $2.6$ & $0.04$ & $3.8$ & 0.19 & 0.45 & 0.36 & 8.6 \\
{\color[HTML]{9932CC} Spawn} & particle spawn & isotropic  & $3.2$ & $0.04$ & $5.6$ & 0.27 & 0.47 & 0.26 & 8.5 \\
\enddata
\end{deluxetable*}

\section{Methods} \label{sec:data}

\subsection{FIRE-2 Model} \label{subsec:fire2}
We analyze cosmological zoom-in simulations of MW mass galaxies ($M_{\rm vir} \sim 10^{12} \,\rm M_{\odot}$) as part of the Feedback in Realistic Environments (FIRE) updated galaxy formation model \citep[FIRE-2][]{Hopkins2014, Hopkins2018}. The core FIRE-2 physics includes radiative gas cooling, star formation in dense, self-gravitating gas, and stellar feedback from supernovae, stellar mass-loss, and radiation that interact with the surrounding gas. Our specific subset of FIRE-2 runs includes magnetohydrodynamics (MHD) and cosmic rays (CRs) from stellar feedback. The cosmic ray diffusion coefficient is $\kappa \approx 3 \times 10^{29} \,\rm cm^{2}\,s^{-1}$ and details on the emission and propagation of CRs are described by \cite{Chan2019}. 

The initial gas particle mass resolution of these simulations ($m_{\rm b} = 5.6 \times 10^{4} \,\rm M_{\odot}$) is slightly higher mass (lower resolution) than the fiducial FIRE-2 MW mass runs, but still captures a realistic, multiphase interstellar medium \citep[see][]{Hopkins2018}. The addition of AGN feedback within the FIRE-2 galaxy formation model, then, allows us to detail interactions between SMBH feedback and the ISM/CGM in real time.  

\subsection{AGN Feedback Model} \label{subsec:agnparams}

Here, we provide a brief summary of our AGN feedback model. For full details, we refer the reader to \citet{Wellons2023} and references therein.  The model connects gas supply near the black hole to SMBH growth, which is then linked to feedback in the form of radiation pressure, mechanical winds, and cosmic rays. The simulations include a black hole ‘particle' that represents both the black hole itself as well as an accretion disk, which has a separate mass reservoir. 

Mass flows from local gas into the accretion disk at a rate 
$\dot{M}_{\rm acc} = \eta_{\rm acc} \times M_{\rm gas} \tau^{-1}$, where $M_{\rm gas}$ is the gas mass within the BH accretion kernel and $\tau$ is a local dynamical time. Here,  $\eta_{\rm acc}$ is an accretion efficiency function, which we model using the gravitational torque approach of \citet{2011hopkinsQuataert}.  In this picture, instabilities in the gravitational potential at various scales of the galaxy transfer angular momentum and drive gas towards the center.  The value of  $\eta_{\rm acc}$ depends on the evolving accretion disk mass and the black hole mass. See \citet{Wellons2023} for details. 

The black hole grows at a rate $\dot{M}_{\rm BH}$ from the accretion disk material following expectations for an $\alpha$-disk model \citep{Shakura1973}. The black hole accretion rate drives all three feedback channels: radiation pressure, mechanical winds, and cosmic rays. 

For radiation pressure, we assume that a fraction $\epsilon = 0.1$ of the mass-energy accreted onto the black hole is converted into radiation, $L_{\rm rad} = \epsilon \times \dot{M}_{\rm BH} c^{2}$ \citep{Shankar2020}.  Radiative transport is tracked with the LEBRON method \citep[][]{Hopkins2020}, and we adopt a template spectrum of a quasi-stellar object (QSO) to calculate photoionization, photoelectric, and Compton heating/cooling rates.  Photon momentum flux ($\dot{p}$) is then calculated using the luminosity absorbed by a gas element ($L_{\rm abs}$). In principle, our model allows for an efficiency parameter ($\eta_{\rm RP}$) to potentially account for unresolved substructure below the resolution limit that could absorb radiation: $\dot{p} = \eta_{\rm RP} \times L_{\rm abs}/c$.  All simulations analyzed in this paper set $\eta_{\rm RP} = 1$, such that the radiation pressure is unmodified from its naive value.

Mechanical feedback accounts for matter that is launched away from the accretion disk as an outflow.  We assume that the kinetic energy rate of the wind is $\dot{E} = \dot{M}_{\rm BH} \times v_{\rm wind}^{2}/ 2$. The free parameter here is the wind speed.  Observed AGN-driven outflows have velocities ranging from 1,000 km s$^{-1}$ to 30,000 km s$^{-1}$ \citep{Cattaneo2009, 2012Faucher}. With this as a guide, we chose $v_{\rm wind} = 10,000$ km s$^{-1}$ for all runs in this paper.  

In addition to mechanical winds, AGN produce cosmic rays (CRs) from the magnetic fields produced by the accretion disk, which accelerate charged particles to relativistic speeds. We model CRs as a relativistic fluid, incorporating diffusion, streaming, catastrophic and Coulomb/ionization losses, adiabatic work, and pressure coupling between the CRs and gas \citep[see][]{Chan2019}. The CR energy injection rate is calculated as: $\dot{E}_{\rm CR} = \eta_{\rm CR} \times \dot{M}_{\rm BH} c^2$. The value of the CR efficiency parameter in our runs is set to $\eta_{\rm CR} = 0.01$.

All of the AGN simulations we present in this paper employ the same model assumptions for determining the rate of energy and momentum injected as feedback. The difference, as we discuss in the next subsection, relates to how we couple that feedback numerically and geometrically within the simulation.

\begin{figure*}[ht!]
\centering
\includegraphics[width=.85\textwidth]{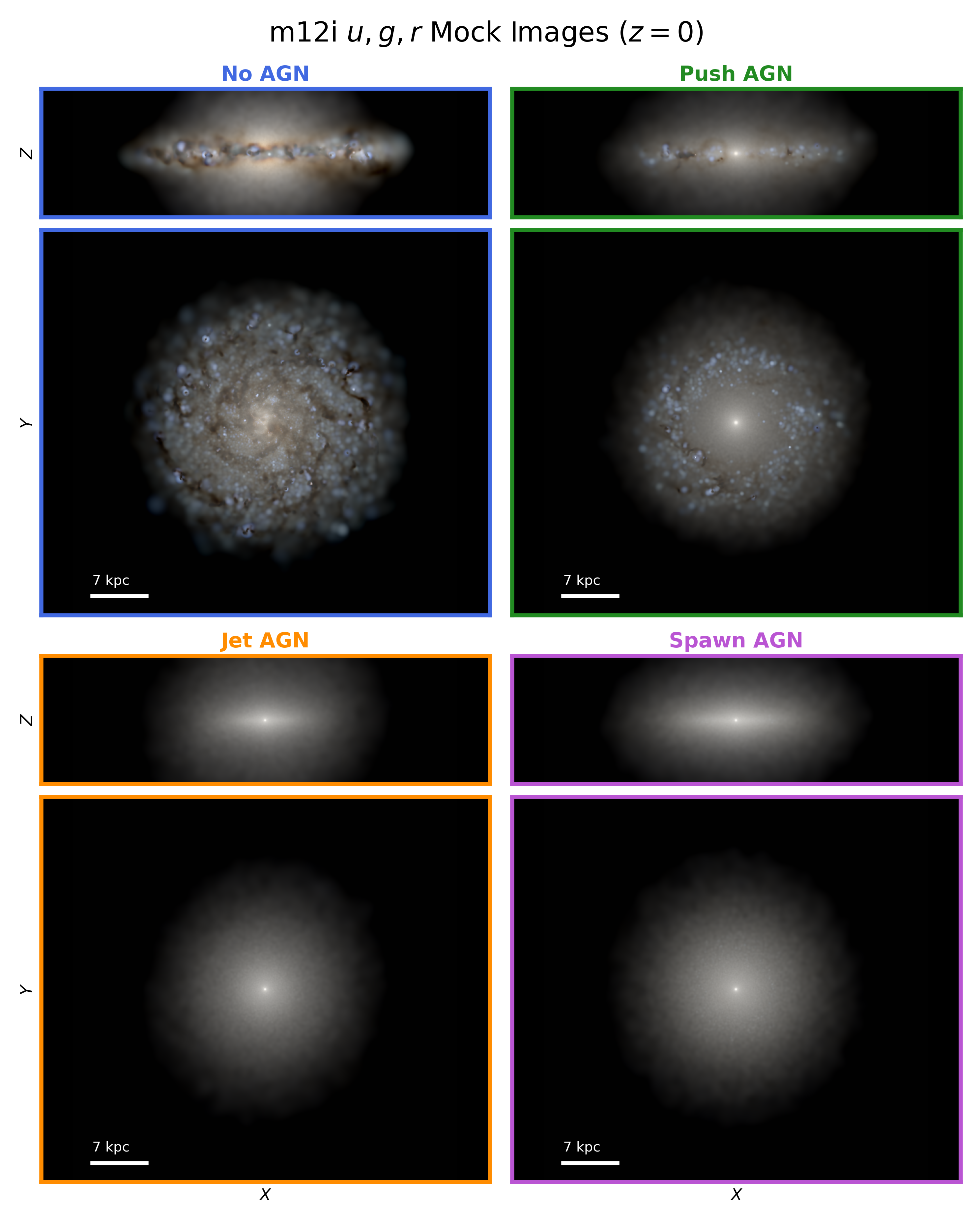}
\caption{SDSS $u$, $g$, $r$ band composite images of the four m12i runs at $z=0$ with a total field of view of 25 kpc. The top row shows face-on and side views of the \small{No AGN} run, outlined in blue, and the \small{Push AGN} run, outlined in green. The second row shows the same views for the \small{Jet AGN} run, outlined in orange, and the \small{Spawn AGN} run, outlined in purple. Without AGN feedback, this halo produces an extended disk with flocculent spiral structures. The \small{Push AGN} implementations produce a less extended disk, with blue spiral features confined to a ring in the outer galaxy, while the inner bulge region is more prominent. The \small{Jet AGN}  and \small{Spawn AGN} runs both produce lenticular morphologies that are redder in color and without prominent star-forming disks.  
\label{fig:mockm12i}}
\end{figure*}

\begin{figure*}[ht]
\centering
\includegraphics[width=\textwidth]{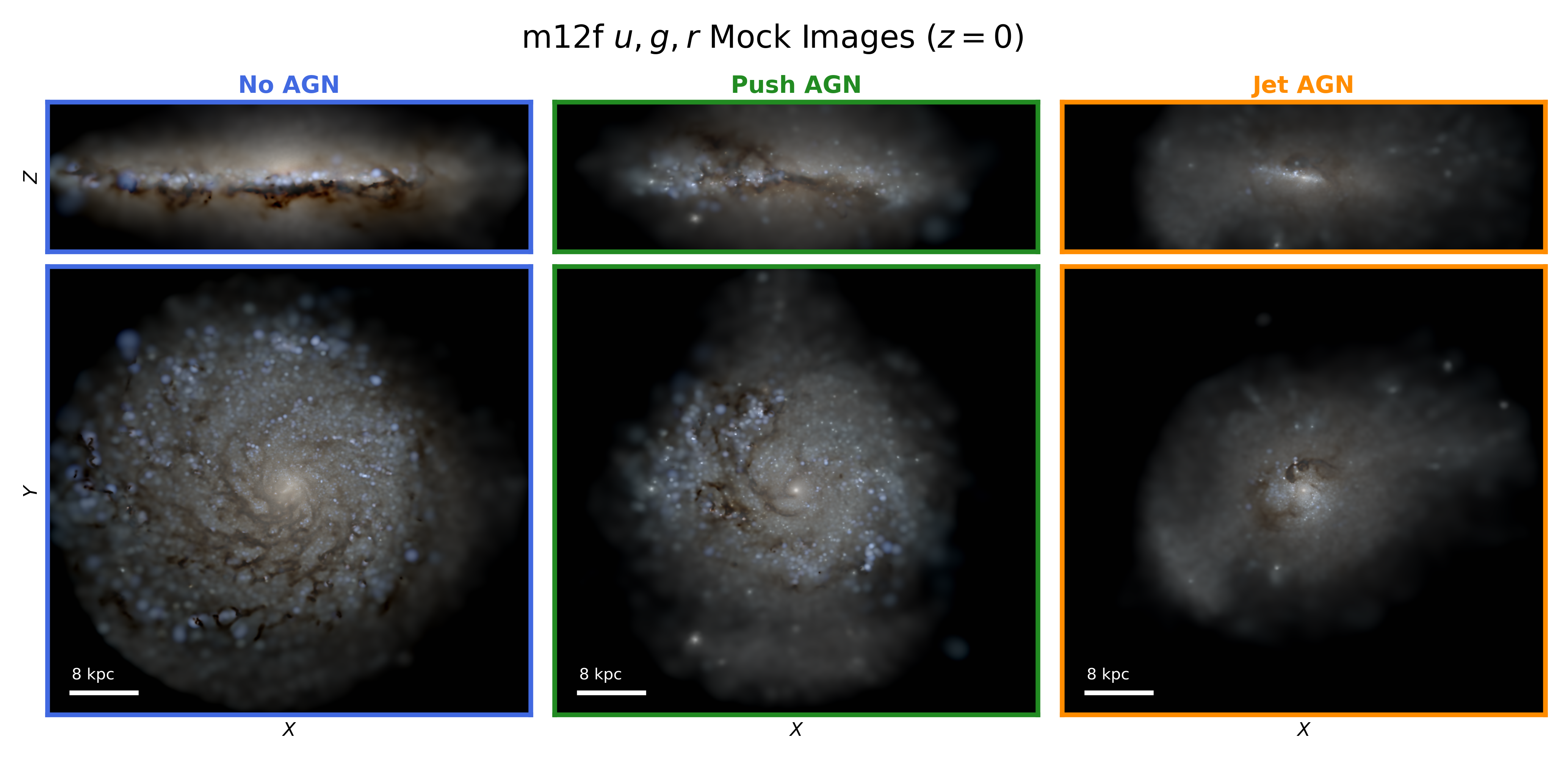}
\caption{SDSS $u$, $g$, $r$ band composite images of the three m12f runs (there is no \small{Spawn AGN} implementation for this galaxy). Each column shows a different AGN implementation with \small{No AGN} outlined in blue, \small{Push AGN} outlined in green, \small{Jet AGN} outlined in orange. Without AGN feedback, this halo produces an extended star-forming disk. The \small{Push AGN} implementation produces a less extended disk, with a more disturbed morphology (owing to a late-time merger).  The \small{Jet AGN}  run lacks an extended star-forming disk but instead shows a disturbed morphology and low-surface brightness features that are indicative of a more significant merger (in a relative sense -- the main progenitor is smaller, owing to effective AGN feedback).   
\label{fig:mockm12f}}
\end{figure*}

\subsection{Simulations} \label{subsec:sims}
We study a subset of seven FIRE-2 simulations with AGN feedback taken from \cite{Wellons2023}, who explore different modeling choices for SMBH accretion and feedback in a suite of cosmological zoom-in simulations across a wide range of halo masses. We chose a small subset of these runs that: a) sit at the mass scale where galaxy morphological bimodality is seen, b)  form prominent star-forming disks at $z=0$ without SMBH feedback, and c) have specific choices of SMBH feedback parameters that, when applied, obey observed galaxy scaling relations. 

There are two sets of initial conditions: ``m12f" and ``m12i", with halo masses at $z=0$ of $1.4 \times 10^{12} \,\rm M_{\odot}$ and $1.0 \times 10^{12} \,\rm M_{\odot}$ respectively. Each of these has a base run without AGN feedback, producing galaxies with star-forming thin disks at $z=0$. The m12i halo has a more quiescent accretion history, while late-time minor mergers are more common in m12f. 

The base runs, labeled \small{``No AGN,''}, each have up to three counterpart runs with AGN feedback implementations labeled \small{``Push AGN,''} \small{``Jet AGN,''}, and \small{``Spawn AGN.''} As previously described, each of these implementations uses identical assumptions for determining the rate of energy and momentum injection from mechanical winds, cosmic rays, and radiation pressure. They differ only in {\em how} the feedback energy and momentum are imparted in the simulation:

\begin{itemize}
    \item \small{\textbf{Push AGN}} deposits the mass, energy, and momentum directly into gas cells within the interaction kernel or radius of interaction in an isotropic fashion around the black hole. \cite{Hopkins2016} previously studied variations of this Push implementation on the impact of AGN winds. 

    \item \small{\textbf{Spawn AGN}} creates new, high-resolution ($\sim100 \,\rm M_{\odot}$) energetic particles in an isotropic sphere around the BH and combines the BH feedback energy and momentum with gas cells along the way. \cite{Torrey2020} explored variations of the \small{Spawn} implementation on the impact of AGN winds in a multi-phase galactic disk. 

    \item \small{\textbf{Jet AGN}} also creates new, high-resolution particles near the BH.  However, they are deposited along a collimated outflow in the $\pm$ z-axis (defined by the angular momentum of the accreted material). \cite{Su2021} explored this \small{Jet} implementation and variations on the model to determine which AGN jet injection can quench massive galactic systems.
\end{itemize}

\subsection{Circularity Parameter} \label{subsec:circparam}

Our kinematic proxy for morphology is the circularity parameter $\epsilon = j_{\rm z}/j_{\rm c}(\rm E)$, which takes the ratio of a particle's angular momentum in the $\hat{z}$ direction to a circular orbit with the same energy \citep[e.g.][]{Abadi2003, Yu2021}. We calculate specific angular momentum for stars or gas using all respective particles within an enclosed sphere of radius 20 kpc, and we calculate the gravitational potential for the specific energy using the total mass within such. An orbital circularity of $\epsilon = 1$ corresponds to a particle on a circular orbit in the plane of a disk defined by the total stellar or gas angular momentum at a given redshift. We employ the morphology demarcations developed by \cite{Yu2021}, which relate a range of orbital circularity values to a morphology. Namely, $1.0 \geq \epsilon > 0.8$ corresponds to a particle classified as \textbf{Thin Disk}; values in the range $0.8 \geq \epsilon > 0.2$ correspond to a \textbf{Thick Disk} classification, and $\epsilon \leq 0.2$ corresponds to a \textbf{Spheroidal} classification. Note that negative values correspond to counter-rotating components. 

These circularity demarcations do not map perfectly onto observationally derived definitions of spheroid, thick, or thin disk components. They instead work as convenient and well-defined kinematic categories that match our qualitative understanding of these components, informed by geometric distinctions \citep[see][]{Sales2012, Benavides2025}. 

\section{Results} \label{sec:results}

\subsection{Visual Morphologies from Mock Images \label{subsec:mockim}}

We begin by presenting mock images illustrating the morphological changes that occur under a given AGN feedback model.  We use an adapted version of FIRE Studio \citep{Hopkins_2005, Gurvich2022} to create mock images, as described in Section 2.2 of \citet{Klein2024}. To derive the luminosity of each star particle, we use a mass-to-light ratio based on the particle's mass, age, and metallicity  \citep{Chabrier2003}. We account for extinction via Thomson scattering, the photoelectric effect, and dust absorption. Each particle is smoothed using a cubic spline kernel with a smoothing length of 1.4 times the particle's gravitational softening length.

Figure \ref{fig:mockm12i} shows SDSS $u$, $g$, and $r$ band composite images for face-on and side views of the four m12i runs.  The field of view of each galaxy is 25 kpc. Clockwise from the upper left, we show \small{No AGN}, \small{Push AGN}, \small{Spawn AGN}, and \small{Jet AGN}. Without AGN feedback, m12i has a prominent thin disk with a flocculent spiral pattern and recent star formation visible in the bright blue light.  Optically thick dust regions are also apparent in the image. The \small{Push AGN} run produces a less extended and thinner disk, with noticeably fewer new stars in the central bulge region. Blue stars are confined mostly to a ring in the outer disk in this run. In the second row, the \small{Jet AGN} and \small{Spawn AGN} runs of m12i produce classic S0/lenticular morphologies, with no young stars present and no obvious dust lanes. \small{Jet AGN} has the smallest radius, with \small{Spawn AGN} slightly more extended but still smaller than either of the runs on the top row. 

Figure \ref{fig:mockm12f} provides similar images for the three m12f runs, with \small{No AGN}, \small{Push AGN}, and \small{Jet AGN} shown left to right. The \small{No AGN} run again produces a galaxy with a prominent thin disk, a flocculent spiral pattern, and significant star-forming (blue) light. Note that this disk is roughly twice the radial size of the \small{No AGN} run of m12i, although they have similar stellar masses. In both AGN models of m12f, we see a reduction of the young stellar population (blue light), a diminished dusty gas reservoir (dark/optically thick regions), and a reduced presence of spiral disk features. The \small{Jet AGN} run (far right) produces a significantly smaller disk compared to the \small{No AGN} run (far left). Moreover, this disk lacks blue, star-forming light.

Particularly striking in the \small{Jet AGN} image of m12f are low-surface brightness features indicative of a recent, gas-rich minor merger. The same late-time gas-rich merger also occurs in \small{Push AGN} and \small{No AGN}, and is likely the origin of the disturbed and lopsided appearance of \small{Push AGN} here. No lasting morphological disturbance occurs for the \small{No AGN} case. As we discuss below, the merger triggers a late-time starburst in \small{Jet AGN} and is responsible for most of the blue light in its image. Interestingly, the same merger occurs slightly earlier in both \small{Push AGN} and \small{No AGN} and has less dramatic effects on their star formation histories.  As we discuss below, the strong reaction to this merger in the \small{Jet AGN} case is likely because the main galaxy in this run has the smallest baryonic mass at the time of the interaction; this means that the merging galaxy is more significant by comparison. 

\begin{figure}[ht!]
\centering
\epsscale{1.1}
\plotone{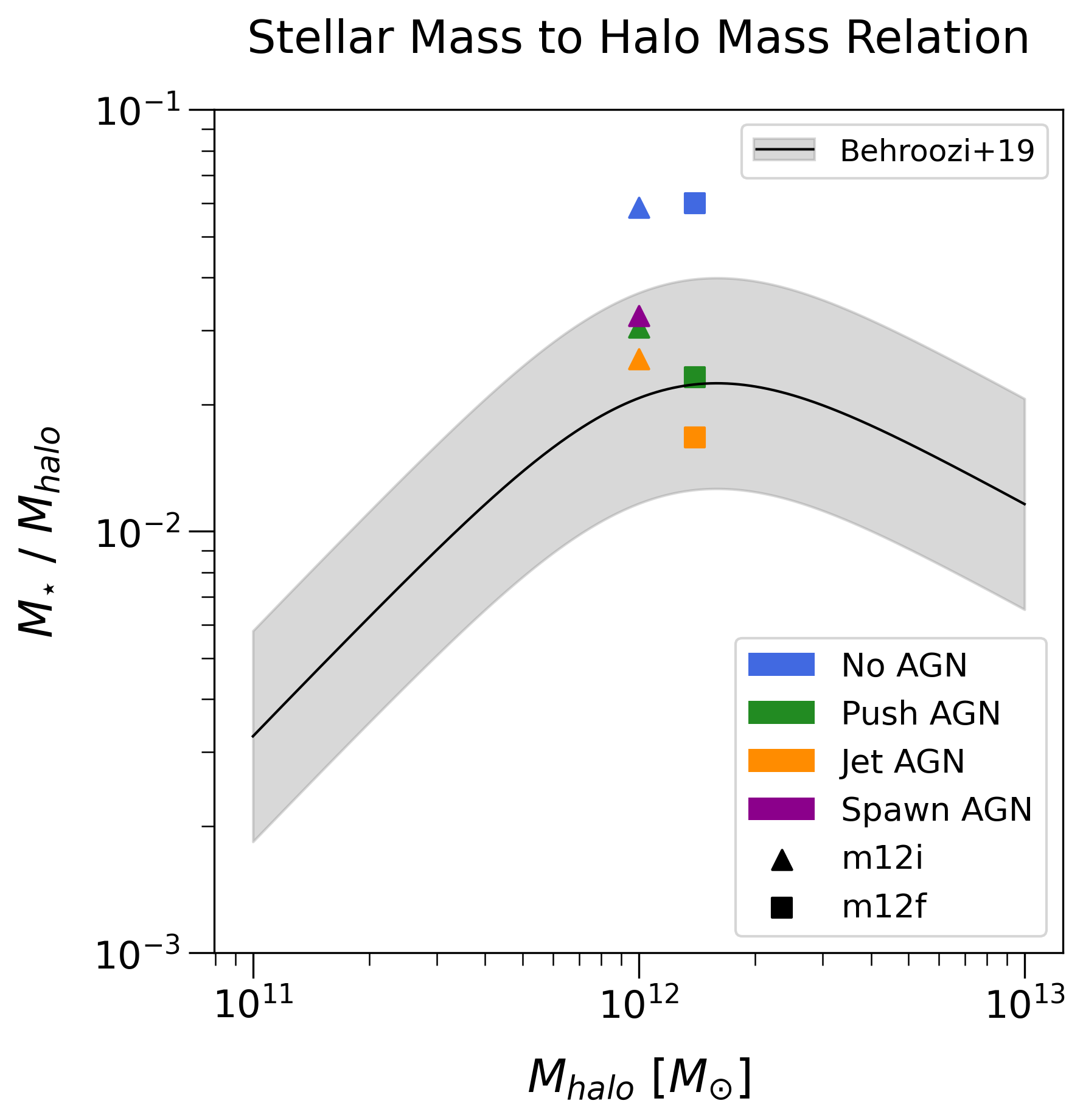}
\caption{Stellar Mass to Halo Mass relation for our seven runs (points). The black line is the observationally constrained relation at $z=0$ adapted from \cite{Behroozi2019}, and the shaded gray region is the 0.25 dex scatter expected for central galaxies in halo masses around $10^{12} \,\rm M_\odot$. We use the stellar mass within 20 kpc. The m12i runs are triangles, and the m12f runs are squares. The colors indicate the feedback model: blue for \small{No AGN}, green for \small{Push AGN}, orange for \small{Jet AGN}, and purple for \small{Spawn AGN}. The runs without AGN feedback land above the observed relation, while those with AGN feedback land within the expected range. Among the three AGN models, the \small{Jet AGN model} is the most effective at suppressing star formation.
\label{fig:SMHMplot}}
\end{figure}

\begin{figure*}[t!]
\centering
\plotone{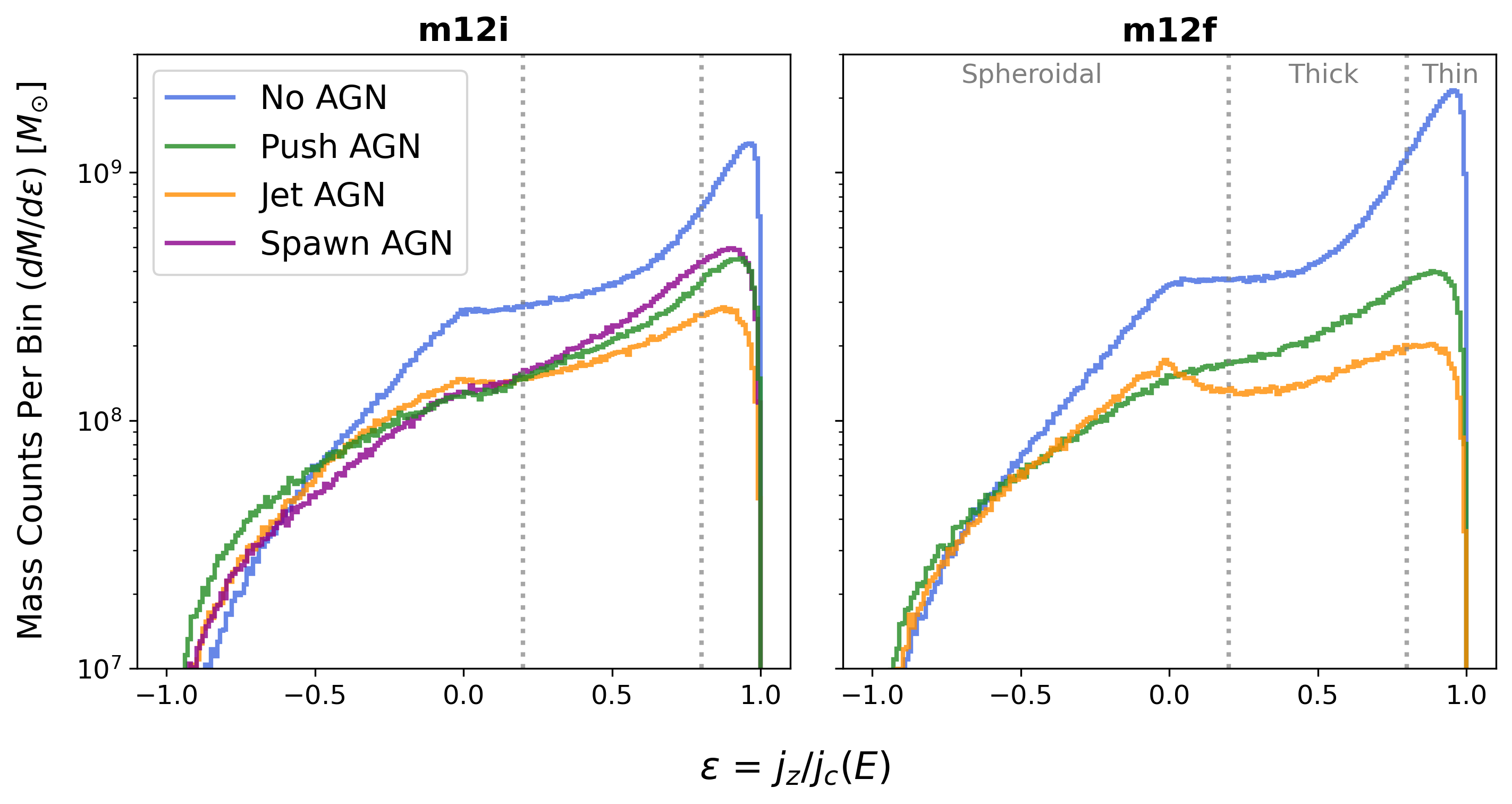}
\caption{Circularity distributions at $z=0$ of the all the stars within 20 kpc for m12i (left) and m12i (right) for runs with \small{No AGN} (blue), \small{Push AGN} (green), \small{Jet AGN} (orange), and \small{Spawn AGN} (purple). An $\epsilon \sim 1.0$ corresponds to a thin disk orbit, $\epsilon \sim 0.5$ is a thick disk orbit, and $\epsilon \sim 0.0$ corresponds to a spheroidal orbit. We see that all AGN feedback models preferentially suppress the abundance of stars at higher circularities. This is seen most clearly in the suppression of thin-disk orbits.  \small{Jet AGN} suppresses star formation most efficiently and most preferentially among thin-disk stars, see Section \ref{subsec:morph}.
\label{fig:circplot}}
\end{figure*}

\subsection{Global Star Formation Suppression}

Figure {\ref{fig:SMHMplot} provides an example of how AGN feedback in our runs affects the overall stellar mass formed in each system.  Each point shows the fraction of total stellar mass to halo virial mass as a function of virial mass for the m12i (triangles) and m12f (squares) runs, compared to the relation required to match galaxy abundances from \citet{Behroozi2019}. The shaded gray region shows 0.25 dex scatter about the median relation, which is the scatter expected for central galaxies in halo masses around $10^{12} \,\rm M_\odot$. The non-AGN runs (blue) sit above the SMHM relation and have systematically higher stellar masses compared to their counterparts with AGN feedback. All AGN runs fall within observed expectations, suppressing overall star formation. The \small{Jet AGN} model produces the lowest total stellar mass in both halos. 

\subsection{Morphological Components at $z=0$} \label{subsec:morph}}

\begin{figure*}[ht!]
\epsscale{1.1}
\centering
\plotone{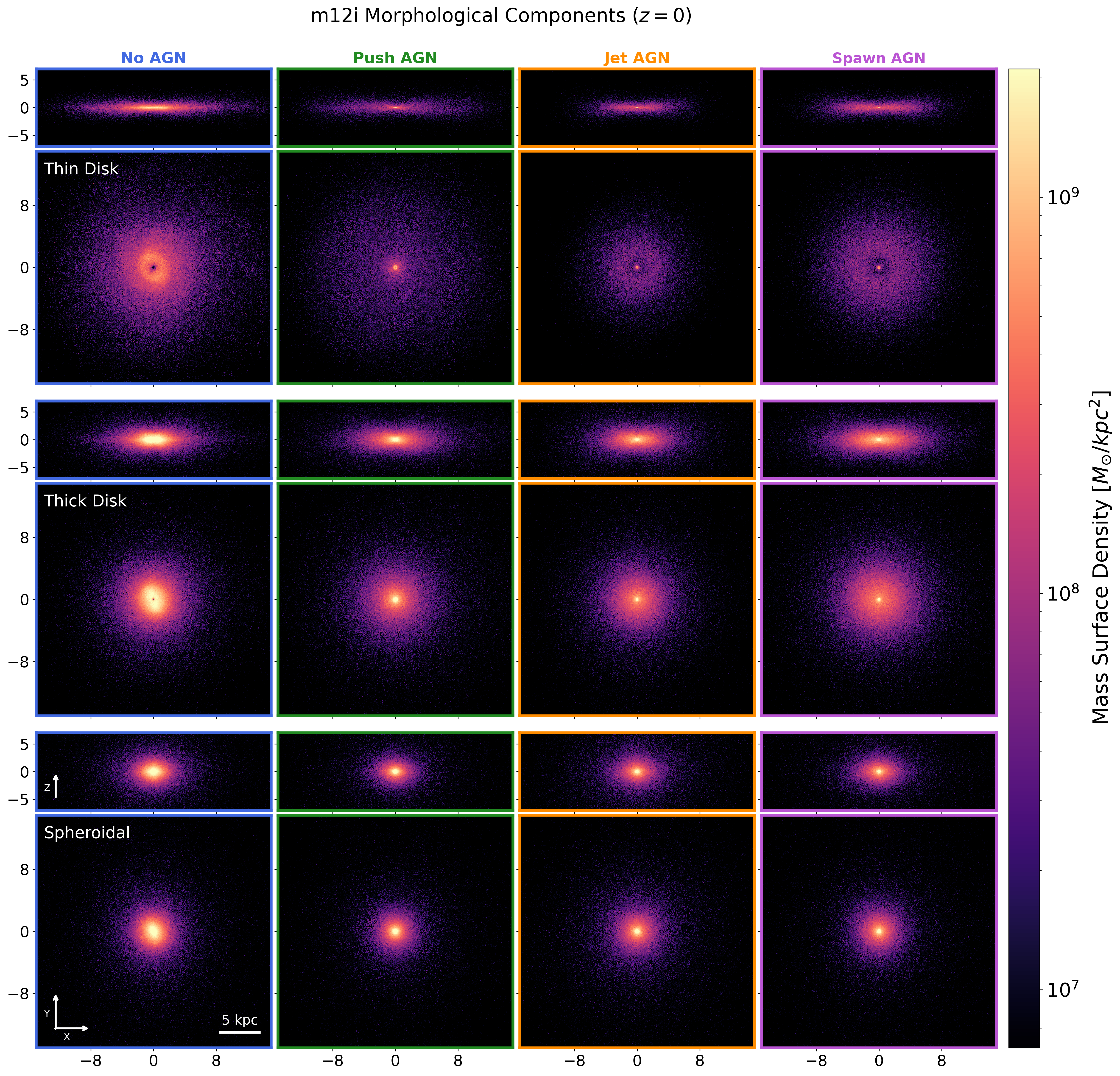}
\caption{Stellar mass weighted 2D histograms of m12i at $z=0$. From left to right, each column of the figure shows the distribution of stars under \small{No AGN}, \small{Push AGN}, \small{Jet AGN}, and \small{Spawn AGN} feedback. The top two rows show face-on and side views of the thin disk component ($\epsilon \geq 0.8$), the middle two rows show the thick disk component ($0.2 \leq \epsilon < 0.8$), and the bottom two rows show the spheroidal component ($\epsilon < 0.2$).  There is a noticeable reduction in mass of the thin disk component for all three AGN runs, with each AGN model differing in the extent/distribution of the thin disk. There are more subtle differences for the thick disk and spheroidal components, namely the lack of bar/boxy structure in all three AGN runs.
\label{fig:2dhist_m12i}}
\end{figure*}

\begin{figure*}[hp!]
\epsscale{1.05}
\centering
\plotone{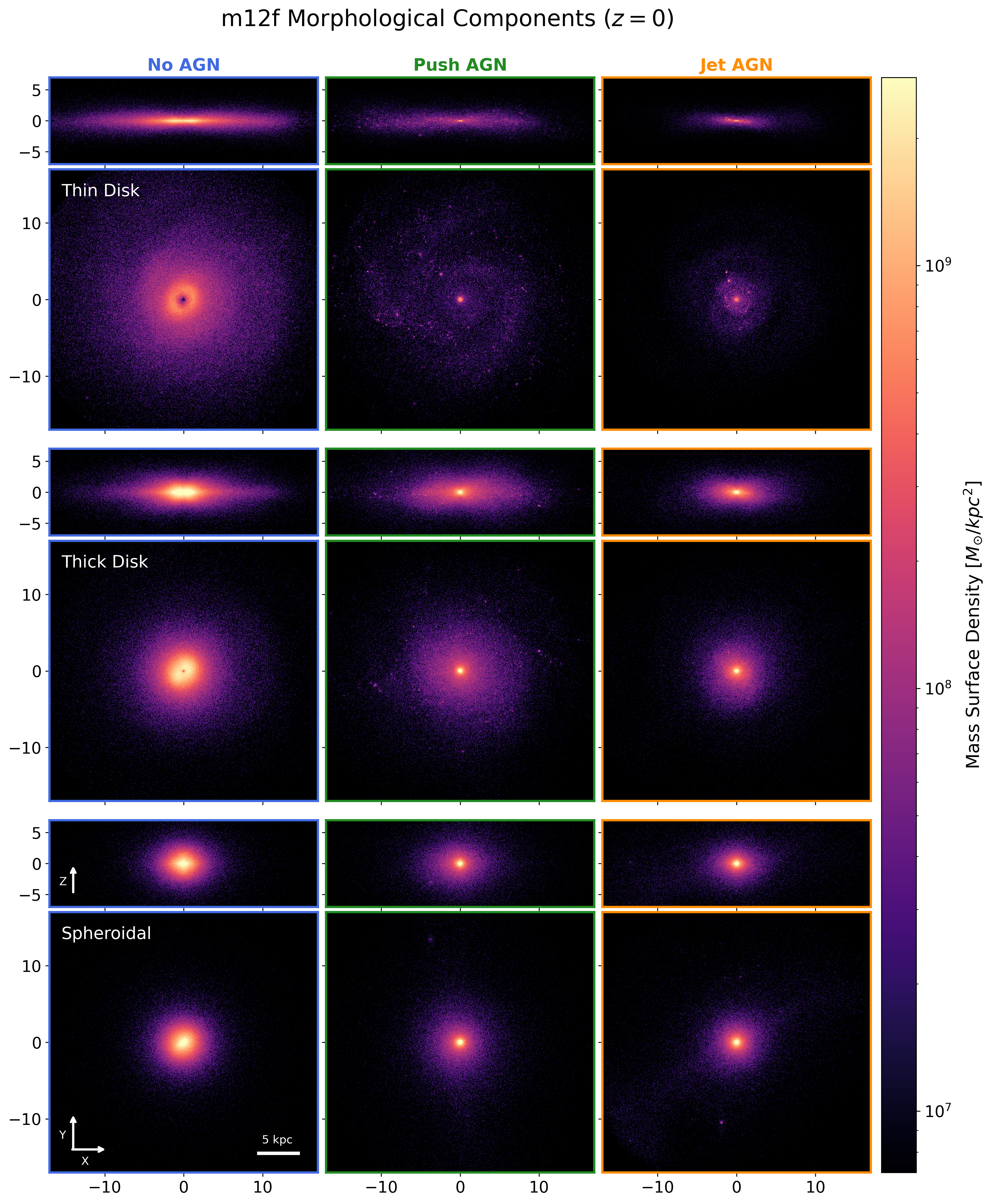}
\caption{Similar to Figure \ref{fig:2dhist_m12i} for m12f. The same broad differences in stellar mass and thin disk distribution occur in m12f, with AGN runs showing an overall reduction in the stellar mass and radial extent of all components, but especially the thin disk component. Under \small{Push AGN}, stars are distributed in the outer regions of the thin disk, while for \small{Jet AGN}, stars are found mostly in the inner region of the thin disk. The thick disk and spheroidal component show subtle differences in the bar/boxy structure between AGN models. 
\label{fig:2dhist_m12f}}
\end{figure*}

Figure \ref{fig:circplot} shows mass-weighted histograms of star particle circularities within 20 kpc for m12f (left) and m12i (right) at $z=0$. The vertical gray dotted lines mark our circularity classifications for the thin disk ($\epsilon > 0.8$), thick disk ($0.2 < \epsilon < 0.8$), and spheroidal ($\epsilon < 0.2$) components. 

As expected from Figure \ref{fig:SMHMplot}, both m12f and m12i have the highest total stellar masses in their \small{No AGN} runs (blue), and the lowest total stellar masses in the \small{Jet AGN} runs (orange). Importantly, Figure \ref{fig:circplot} reveals that the suppression in star formation among the different AGN runs is not constant with final circularity at $z$=0. All AGN feedback models preferentially suppress the abundance of stars at high circularity compared to runs without AGN feedback. 

The largest differences are seen among thin-disk stars. As listed in Table \ref{tab:one}, without AGN Feedback, these stars make up 40\% and 35\% of the total stellar mass in m12f and m12i, respectively. These fractions drop to 24\% and 26\% for the \small{Push AGN} runs, and further to 16\% and 19\% in the \small{Jet AGN} runs. For \small{Spawn AGN}, the thin disk fraction is 27\% of the total stellar mass of m12i, similar to \small{Push AGN}. 

The thick disk mass fractions are more stable from run-to-run, with only slight increases in runs with AGN.  In m12f, the \small{No AGN} thick disk fraction is 38\%.  It increases a bit to 45\% in \small{Push AGN} and 42\% in \small{Jet AGN}. For m12i, the change in thick disk fraction is similar: from 35\% in the \small{No AGN} to 44\%, 45\%, and 47\% in the \small{Push AGN}, \small{Jet AGN}, and \small{Spawn AGN} runs, respectively (see the Morphology columns in Table \ref{tab:one}).

The spheroidal fractions also show marked increases as we compare non-AGN runs to AGN runs. In m12f and m12i, the \small{No AGN} spheroidal mass fraction is 22\% and 24\% respectively.  For comparison, the \small{Push AGN} versions have higher spheroidal fractions of 31\% and 30\%, and the \small{Jet AGN} runs rise further to 43\% and 36\%, respectively.  

Perhaps the clearest way of characterizing the morphological differences between non-AGN and AGN runs is by looking at the ratio of spheroidal mass to thin-disk mass (S/T).  As can be inferred from the fractions listed in Table \ref{tab:one}, in m12f, the ratio of spheroidal mass to thin-disk mass is $\rm S/T= 0.55$ for the \small{No AGN} run. In \small{Push AGN}, the ratio rises to $\rm S/T=1.3$ and further to $2.7$ in \small{Jet AGN}. In m12i, $\rm S/T=0.68$ for the non-AGN run.  The ratio rises to $1.15$, $1.89$, and $0.96$ for \small{Push AGN}, \small{Jet AGN}, and \small{Spawn AGN}, respectively.

Figures \ref{fig:2dhist_m12i} and \ref{fig:2dhist_m12f} present 2D histograms of the spatial distribution of star particles in m12i and m12f, respectively, divided among the three circularity-defined morphological classifications. Counts are weighted by stellar mass, and the color bar is logarithmic. The top, middle, and bottom rows of the figures visualize the thin disk, thick disk, and spheroidal components for each of the AGN models of the indicated halo. 

Starting with the top row of Figure \ref{fig:2dhist_m12i}, we see that the mass-weighted thin disk component of m12i changes dramatically from run to run. In the \small{No AGN} run, the thin disk is prominent and spans $\sim 25$ kpc.  Interestingly, in the face-on view, we see a hole in the middle, because none of the stars in the center of this galaxy have very circular orbits. Moving right, the thin disk component in the \small{Push AGN} run is more reduced. Though only slightly smaller in overall physical extent (spanning $\sim 20$ kpc), the surface density is much lower, with most of the mass concentrated at a very small radius, where the presence of the black hole may have created conditions for more circular orbits among stars near the center compared to the run without AGN feedback. In contrast, the thin disk component spans only $\sim 10$ kpc in the \small{Jet AGN} run, where the surface density is reduced compared to the run without AGN, but is more uniform with radius than in the \small{Push AGN} case.  The thin disk of the \small{Spawn AGN} case resembles a larger version of its \small{Push AGN} counterpart, with a diameter of $\sim 15$ kpc.

The second and third rows of Figure \ref{fig:2dhist_m12i} show that the thick disk and spheroidal components of m12i are more similar from run to run than we saw for their thin disks.  The main difference is that both the thick disk and spheroidal components of the \small{No AGN} run display bar-like central light concentrations; these are absent in the three AGN runs. 
The radial extent of the thick disks is roughly the same in all cases ($\sim 15$ kpc diameter), except for the \small{Jet AGN} run, where the thick disk is about the same size as its thin disk component in the same run ($\sim 10$ kpc diameter). The spheroidal components are almost identical in size across four runs, except the aforementioned bar-like central feature in the \small{No AGN} case.

Figure \ref{fig:2dhist_m12f} shows qualitatively similar trends. As with m12i, the physical differences are most apparent among the thin disk components (top row).  The thin disk of m12f is very extended, with a diameter of nearly 40 kpc. As with m12i, the thin disk component of the \small{No AGN} run of m12f is devoid of stars in the middle, betraying the lack of stars in the center of this galaxy with very circular orbits ({\em not} a physical hole). The thin disks in the AGN runs have much lower surface density, except, again, in the center, where the black hole has eliminated bar-like orbits and enabled more circular orbits at small radii. 

Additionally, there seems to be a difference in the clumpiness of the thin disk component. The \small{Push AGN} model, in particular, shows several stellar clumps among thin disk stars, possibly a result of recent star formation triggered by a late-time gas-rich merger (see below). Similar clumps exist in the \small{Jet AGN} case.  These clumps are visible in the runs with AGN because, in both cases, the late-time triggered star formation represents a higher fraction of the overall thin-disk mass in these runs, allowing recently formed clumps to stand out in a mass-weighted map. As we saw with m12i, the physical extent of the m12f thin disk component in \small{Jet AGN} is the smallest: $\sim 10$ kpc, only one fourth its size in the \small{No AGN} case. 

In the middle row of Figure \ref{fig:2dhist_m12f}, we see that stars classified kinematically as thick disk display an elongated bar-like structure in the \small{No AGN} run. This feature is absent in the two runs without AGN; instead, we see high central mass concentrations when AGN feedback is implemented. The diameter of the thick disk components also decreases from $\sim 30$ kpc to $\sim 25$ kpc to $\sim 10$ kpc as we move left to right from \small{No} to \small{Push} to \small{Jet AGN}. The spheroidal components of m12i (bottom row) are more similar from run to run, except again for the bar-like feature in the \small{No AGN} case, which is not present in the runs with AGN feedback.

\begin{figure*}[ht!]
\centering
\includegraphics[width=\textwidth]{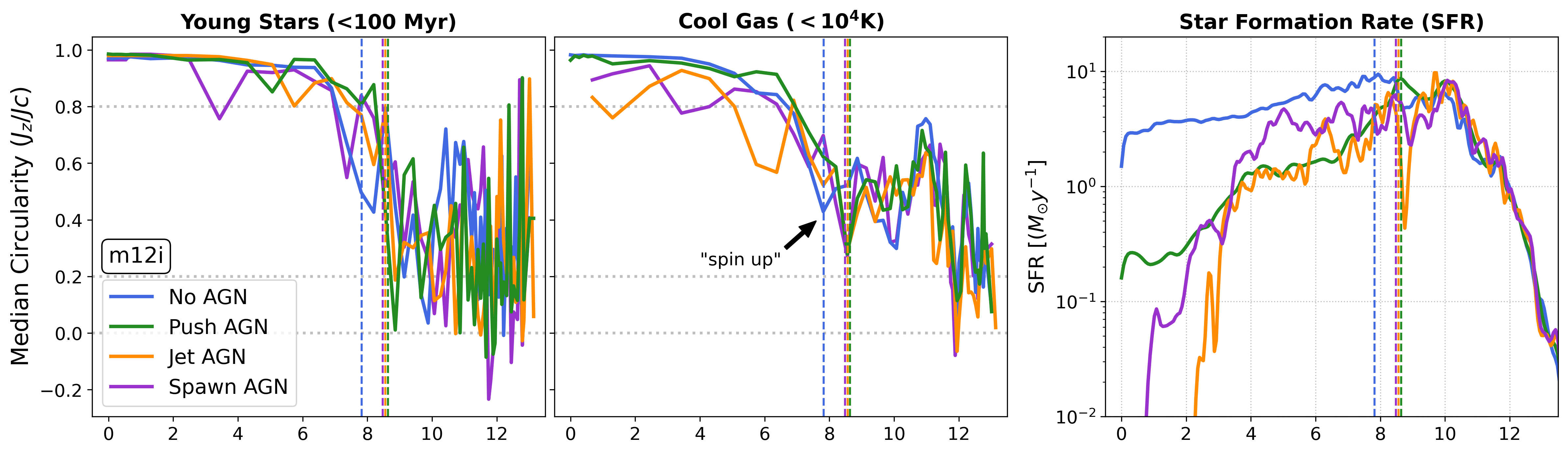}
\includegraphics[width=\textwidth]{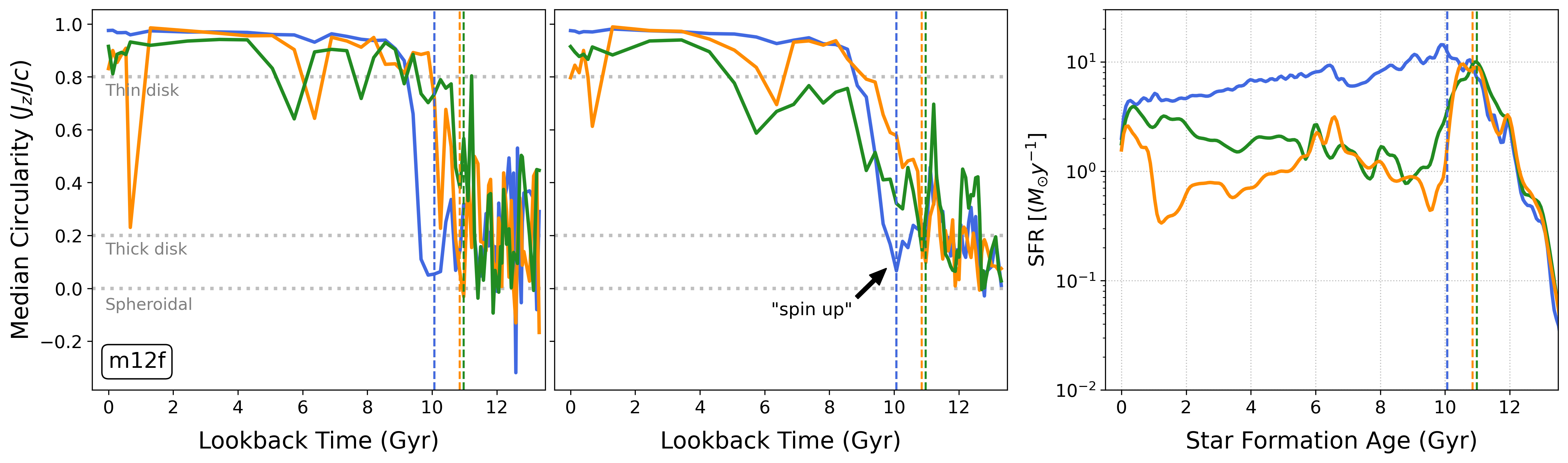}
\caption{The following figure shows star formation suppression relative to spin-up. The left and middle columns show median circularity versus lookback time for young stars ( $t_{\rm age} < 100 \,\rm Myr$) and cold gas ($\rm T < 10^{4} \,\rm K$); the m12i runs are on the top row, and the m12f runs are on the bottom. The \small{No AGN} runs are in blue, \small{Push AGN} runs are in green, \small{Jet AGN} runs are in orange, and the \small{Spawn AGN} run is in purple. Our circularity demarcations for thin disk, thick disk, and spheroidal morphologies are marked with horizontal gray dotted lines. The vertical dashed lines show the ``spin-up" time for all AGN implementations, which signifies the transition from disordered kinematics to a sustained rise towards coherent spin, as discussed in the text. The right column shows the star-formation rate (SFR) as a function of lookback time for the same runs. We see that, in all cases, AGN feedback suppresses star formation relative to the \small{No AGN} runs only after spin-up. The left column shows that the stars that do form in the AGN runs tend to have thin-disk-like orbits at late times. However, the star formation rate is so much lower on average such that the relative fraction of thin-disk stars is suppressed compared to the non-AGN runs. 
\label{fig:circvslbt}}
\end{figure*}

\begin{figure*}[ht!]
\centering
\includegraphics[width=\textwidth]{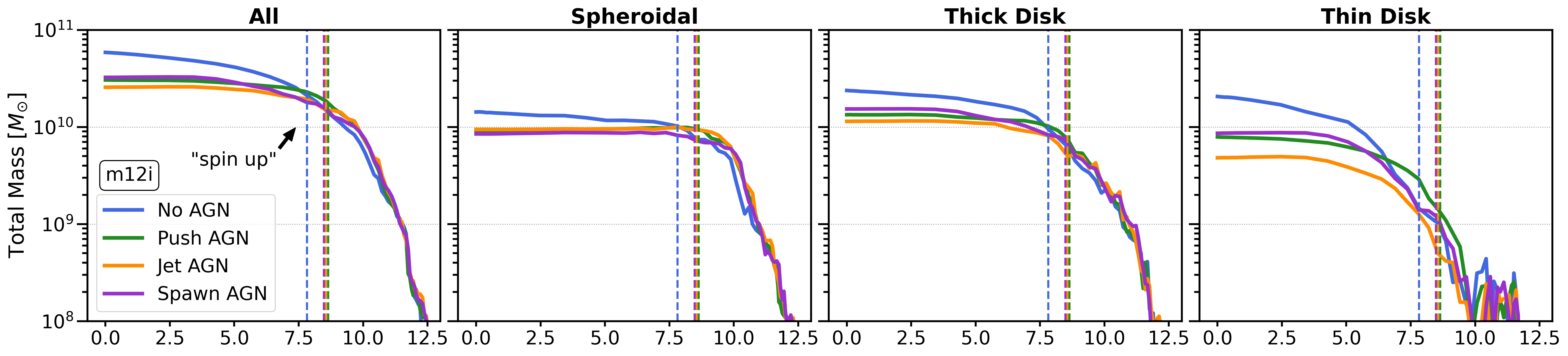}
\includegraphics[width=\textwidth]{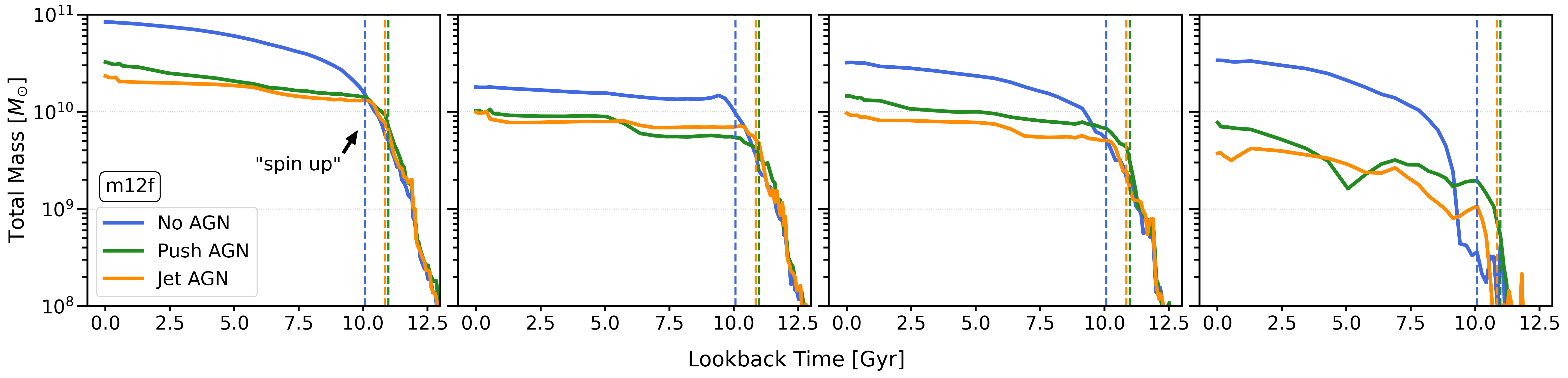}
\caption{The stellar mass within 20 kpc classified within each of our three kinematically-defined components (thin disk, thick disk, spheroidal) as well as the total stellar mass as a function of lookback time. The m12i runs are shown on top, and the m12f runs are on the bottom. The vertical dashed lines show the ``spin-up" time for all AGN implementations.
\label{fig:morph_massplot}}
\end{figure*}

\subsection{Time Evolution \label{subsec:morph_overtime}}

As we saw in the previous subsection, our runs with AGN feedback produce significantly reduced thin-disk components compared to runs without AGN feedback. In this subsection, we investigate the evolution of kinematically identified morphological components over time to explore the origin of the difference.

Figure \ref{fig:circvslbt} displays, for each run, the median circularity versus lookback time for young stars with ages less than 100 Myr old (left) and for cool gas ($\rm T < 10^4 \,\rm K$, middle) as a function of lookback time.  The right panel shows the evolution of the star formation rate\footnote{Unlike the circularity plots, the SFR shown is calculated `archaeologically' from the $z=0$ stellar populations with Gaussian smoothing over 100 Myr timescales. Direct lookback time measurements provide similar results.} (SFR). The m12i and m12f runs are presented in the top and bottom rows, respectively. In these measurements, we focus on material within the central 20 kpc of each galaxy, though we find that the trends we report here are not sensitive to the precise central region we chose. 

Starting with the circularity evolution in the \small{No AGN} runs (in blue, as indicated in the legend), we see that in both m12i and m12f the cool gas and young stars progress through the three kinematic phases as expected (see Section \ref{sec:intro}).  Specifically, in each case, there is an early bursty phase with fairly radial and highly variant circularities in both young stars and cool gas.  This is followed by a ``spin-up" time, after which we see coherently rising circularity with time. Finally, stars and gas enter their ``thin disk" phase, when even the median circularity is very close to unity at late times. The spin-up times for AGN runs are marked by dashed vertical lines in all panels. We follow the spin-up definitions used by \citet{Chandra2024, Myrtaj2026} when the median cool gas circularity begins to rise continuously towards unity. Note that for m12i, the spin-up times of \small{Push AGN}, \small{Jet AGN}, and \small{Spawn AGN} are similar (see Table \ref{tab:one}).  They have been slightly adjusted horizontally in the figure for visibility.

Given the differences in final morphologies in the runs with AGN feedback, it is remarkable that Figure \ref{fig:circvslbt} shows qualitatively similar trends in circularity evolution in the runs with AGN. Specifically, at late times, young stars tend to form on thin-disk orbits across all AGN runs, though the details differ. The median circularities at late times occasionally dip toward the thick-disk regime. Cold gas circularities in the AGN runs deviate more from the non-AGN runs than do the young star circularities. This suggests that, while the gas dynamics are affected by AGN feedback, stars are mostly forming in the fraction of the gas that maintains thin-disk-like kinematics.

One slight difference, which is nevertheless systematic, is that the spin-up times for the AGN runs appear to occur $\sim 0.8$ Gyr {\em earlier} than they do in the non-AGN runs. (See Table \ref{tab:one} for the spin-up times of all runs in this analysis.) The reason for this is unclear, but it may have to do with the role the black hole could play in stabilizing the center-of-mass motion, which seems to precede spin-up in Milky Way mass systems \citep{Myrtaj2026}.

Other observations of note in Figure \ref{fig:circvslbt} include the large downward spike in young-star circularity at late times in the \small{Jet AGN} run of m12f (orange line, bottom left). This coincides with triggered star formation from a late-time gas-rich merger.  We also note that the cold gas content of the \small{Jet AGN} and \small{Spawn AGN} runs in m12i is negligible at late times, which is why those lines are truncated in the top middle panel at a lookback time of 1 Gyr.

With these second-order differences aside, the fact that late-time star formation occurs on {\em {mostly}} thin-disk orbits in the runs with AGN feedback is surprising given that the thin-disk fractions are so reduced in those runs (see Figure \ref{fig:circplot}).  The explanation lies in the SFR panels along the right. Specifically, we see that before spin-up (vertical dashed lines), the SFR for both AGN and non-AGN runs is similar. Only after spin-up does the star formation rate become significantly suppressed in the runs with AGN feedback compared to the non-AGN runs.

The main reason the final morphologies differ in the AGN runs is that star formation is preferentially suppressed during thin disk formation. The stars that {\em do} manage to form in the AGN runs have mostly thin-disk-like orbits.  The fact that there are {\em fewer of them} means that the overall thin disk fraction remains low.  This helps connect the final morphology directly to the impact of star formation suppression driven by AGN feedback: \textit{most late-time star formation occurs in thin disks, but late-time star formation is preferentially suppressed, and this makes the disk less prominent}.

Figure \ref{fig:morph_massplot} shows, from left to right, how the stellar masses in the thin disk, thick disk, spheroidal, and total mass components evolve as a function of lookback time. The m12i runs are shown in the top row, and the m12f runs are shown in the bottom row.  As before, we are measuring the total mass within 20 kpc of the center of each galaxy over time and we classify the stellar mass in each component at the lookback time plotted.

Starting at the left-most column of Figure \ref{fig:morph_massplot}, we see that the total stellar mass evolution in the AGN runs and non-AGN runs begins to deviate in both cases around the time of spin-up (as denoted by the vertical dashed lines). This is to be expected from our earlier discussion of the SFR evolution shown in the right column of Figure \ref{fig:circvslbt}.

The second column from the left of Figure \ref{fig:morph_massplot} shows that the mass classified as belonging to the spheroidal component over time.  We see that spheroidal mass builds up early and stagnates at late times. This helps explain the similarities in spheroidal components of m12f and m12i size across AGN models at $z=0$ seen in Figures \ref{fig:2dhist_m12i} and \ref{fig:2dhist_m12f}. This behavior is expected from what we saw in Figure \ref{fig:circvslbt}: young-star circularities tend to become disk-like at late times. Interestingly, the spheroidal component stops growing slightly earlier in the two AGN runs of m12f, possibly related to the earlier spin-up times in those systems, as well as the drastic suppression in star formation soon after spin-up in the m12f runs.

The differences in evolutionary paths are somewhat more significant in the thick disk components (third column from the left in Figure \ref{fig:morph_massplot}) and become most dramatic in the thin disk components (rightmost column). In keeping with our observations that the AGN runs tend to spin up earlier, the thin disk masses are actually larger in the AGN runs at very early times in most cases, and especially so in m12f, but stop growing in the AGN runs at late times as star formation suppression from AGN feedback becomes significant.

\begin{figure*}[ht!]
\centering
\includegraphics[width=\textwidth]{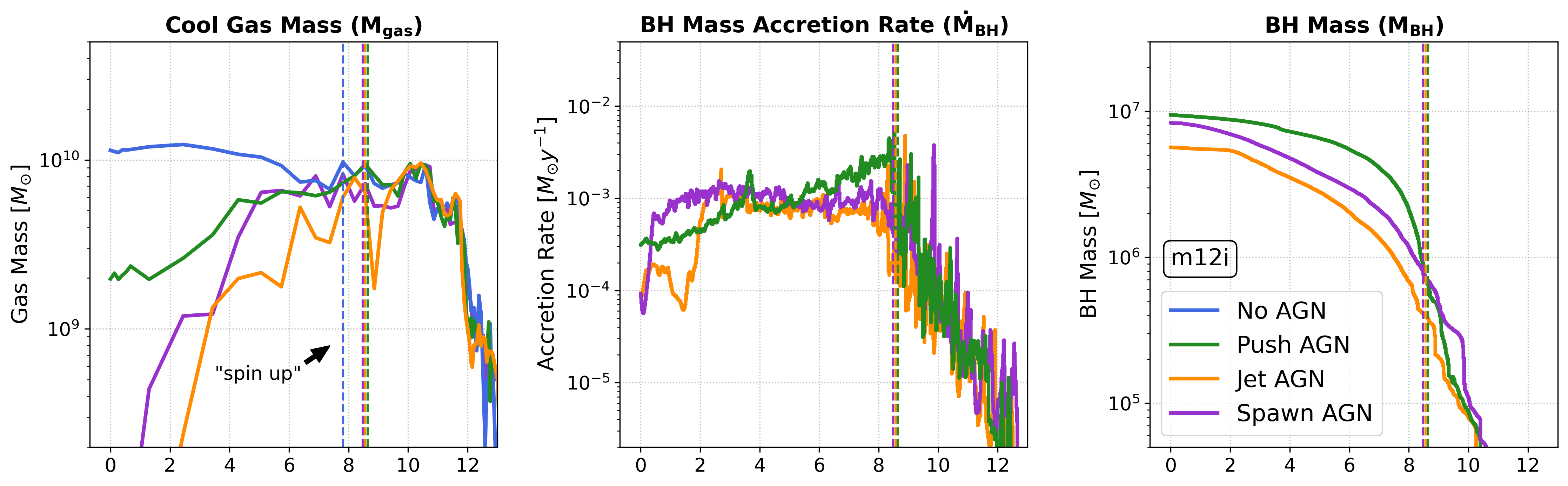}
\includegraphics[width=\textwidth]{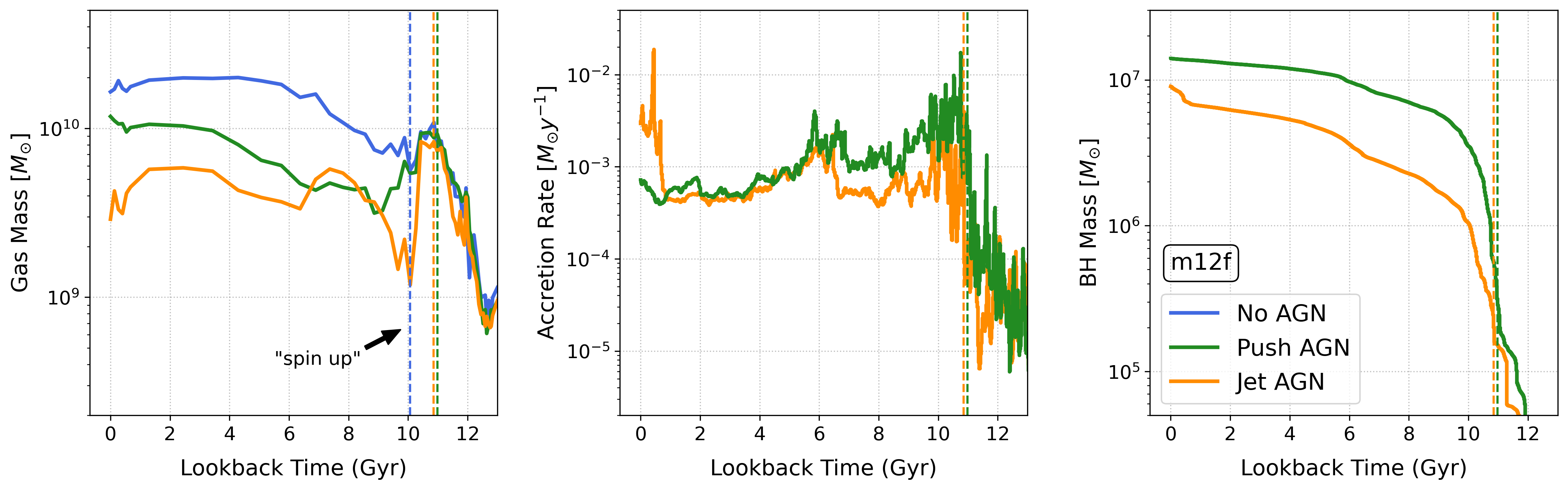}
\caption{The total cool ($<10^{4}$ K) gas mass within 20 kpc, SMBH accretion rate, and SMBH mass versus lookback time for m12f (top) and m12i (bottom). In the vertical dashed lines  we mark the ``spin up" time of the \small{No AGN} runs of m12f and m12i (see Figure \ref{fig:circvslbt} and Section \ref{subsec:morph_overtime} for more details)}
\label{fig:massplot}
\end{figure*}

\subsection{Gas and Black Hole Evolution \label{subsec:mass}}

 Figure \ref{fig:massplot} presents the total cool (T$<10^4 \,\rm K$) gas mass within 20 kpc (left), SMBH mass accretion rate (middle), and the SMBH mass (right) as a function of lookback time for all simulations. As was the case for Figures \ref{fig:circvslbt} and \ref{fig:morph_massplot}, the m12i runs are shown in the top row and the m12f runs are on the bottom row.

Mirroring what we saw with the SFR evolution (right column of Figure \ref{fig:circvslbt}), the total cool gas mass within the galaxy ISM (left panel of Figure \ref{fig:massplot}) remains similar in the AGN and non-AGN runs before spin-up (vertical dashed lines).  After spin-up, the cool gas mass is systematically lower in the AGN runs, which is consistent with the idea that AGN feedback suppresses star formation by suppressing the fuel available for stars to form. The Jet AGN runs, in particular, show dramatic drops in cool ISM gas around the time of spin-up, suggesting that gas is being expelled rapidly when the BH accretion rate peaks (as opposed to slowly starved). In Appendix \ref{app:ISM_CGM} we show the evolution of cool and warm/hot gas in both the ISM and CGM of these galaxies over time, and demonstrate that the main differences are seen in the cool ISM.

As might be expected, the relative differences in cool ISM gas mass between the AGN models roughly track the relative differences in SFR. That is, the \small{Jet AGN} runs have a more pronounced suppression in cool ISM gas mass than the \small{Push AGN} runs, just as in the SFR suppression. Interestingly, the gas mass evolution in the \small{Spawn AGN} run tracks more closely to the \small{Push AGN} case at early times and closer to the \small{Jet AGN} case at late times. That same trend is less apparent in the SFR evolution (Figure \ref{fig:circvslbt}), suggesting that the star-forming gas and cool gas are affected slightly differently at early times in this run. 

The spin-up time also coincides with an interesting feature in the SMBH growth rate evolution (middle panel). We see that $\dot{M}_{\rm BH}$ rises steadily in all cases before spin-up. Just before spin-up, the mass accretion rate tends to spike and then level off.  This is perhaps not surprising, as the gas (fuel) content becomes suppressed after this point and feedback from the black hole begins to limit its own growth. In particular, in both the \small{Jet AGN} and \small{Spawn AGN} runs of m12i, $\dot{M}_{\rm BH}$ plummets at late times as the gas supply drops towards zero. 
In contrast, $\dot{M}_{\rm BH}$ either stays steady or spikes at late times and around 6 Gyr in both AGN runs of m12f, as mergers deliver fresh gas. The Jet AGN run of m12f, in particular, responds to that late-time gas-rich merger with a very large spike in BH accretion rate and an associated dip in the cool ISM mass, as gas is expelled from the galaxy as a result of rapid AGN feedback {\em and} an accompanying burst in star formation (see Figure \ref{fig:circvslbt}).  The same merger in Push AGN does not trigger a spike in BH accretion or a large burst in star formation because it is smaller {\em in proportion to} the main galaxy.  That is, the same merger is a more minor merger in this case because the primary galaxy is larger.  We do see evidence for this merger as a small rise in cool ISM mass near the present day in the \small{Push AGN} run of m12f.

The total black hole mass, which tracks the cumulative impact of all of these processes, ends up being least massive in the runs where gas suppression (and star formation suppression) are the largest.  This makes sense because the fuel supply for black hole growth is most limited in these cases. 

For the specific instance of m12i, which has a fairly quiescent merger history, under AGN feedback, the gas mass decreases continuously as the SMBH mass ($M_{\rm BH}$) increases. Without AGN feedback, the cool gas mass within 20 kpc stays relatively steady, reflecting an equilibrium between star formation and fresh gas accretion. For the more active, merger-filled m12f run, the gas mass within 20 kpc does not decrease continuously in the AGN runs.  Rather, it increases continuously under \small{Push AGN} after an initial outburst, though not as significantly as the \small{No AGN} run, and the \small{Jet AGN} run stays relatively stagnant in its local gas mass. The difference for m12f is the presence of more gas-rich mergers throughout m12f’s evolution, which give rise to its growing gas mass within 20 kpc over cosmic time, larger overall stellar mass, and a more extended disk without AGN feedback.

\section{Discussion} \label{sec:discussion}

\begin{figure*}[hp!]
\centering
\includegraphics[width=\textwidth]{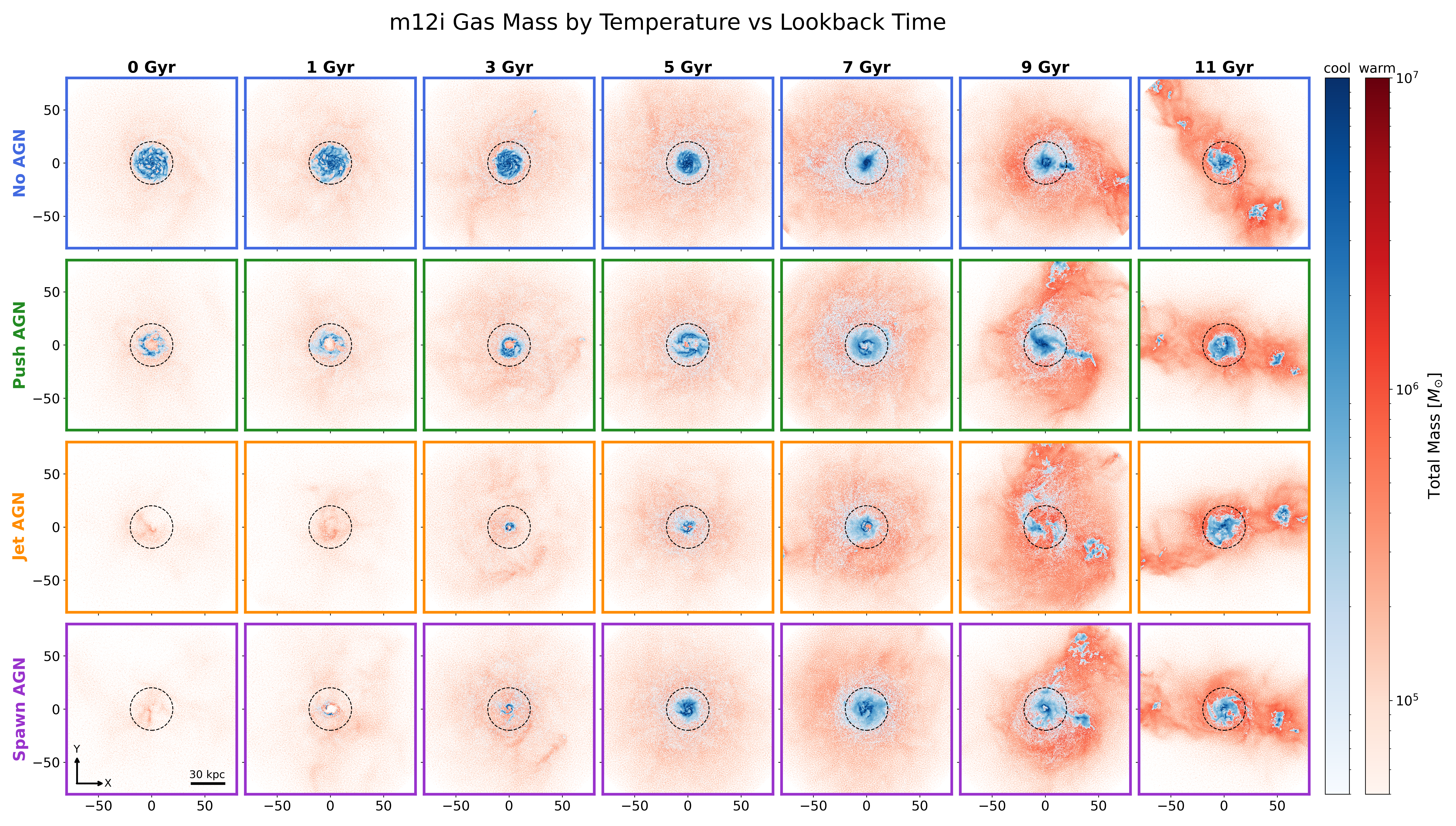}
\includegraphics[width=\textwidth]{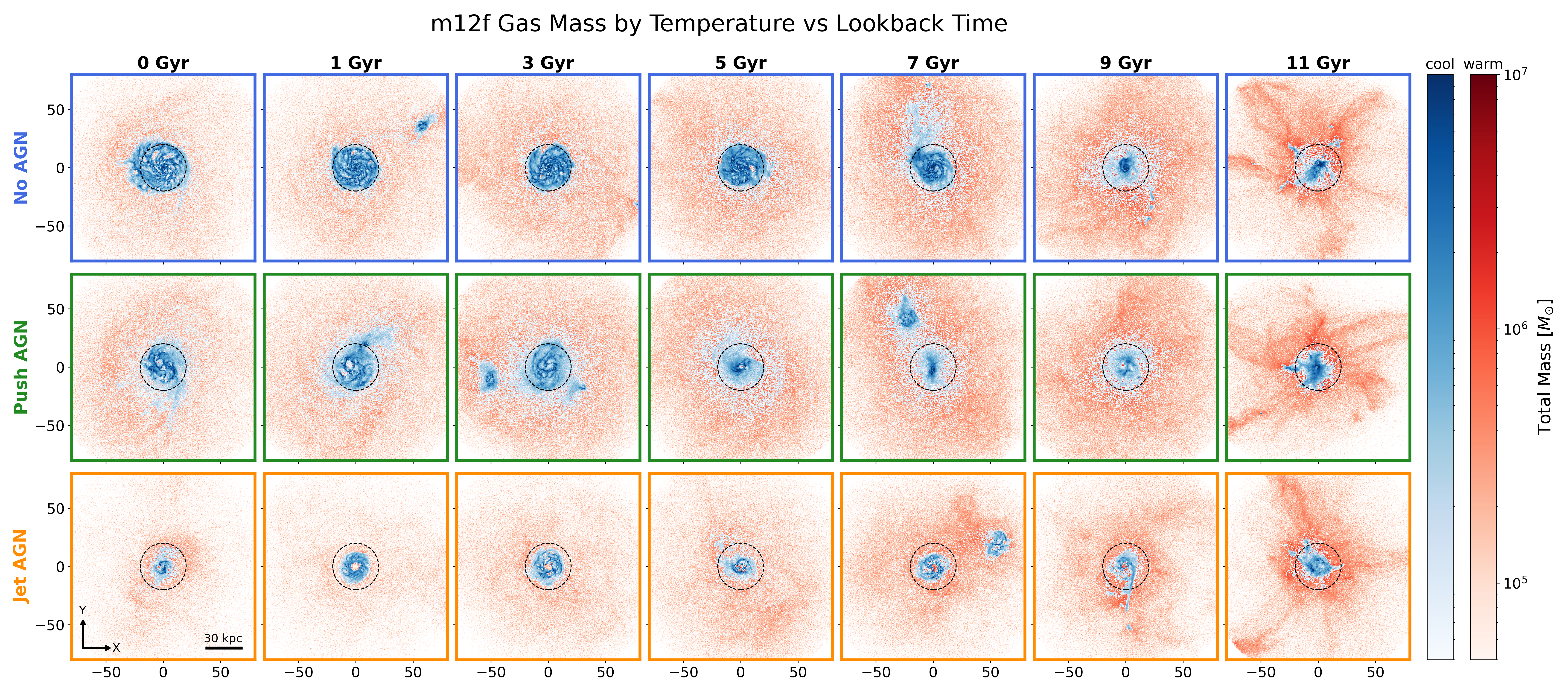}
\caption{2D histograms of gas mass split by temperature showing face-on images of hot gas ($T > 10^{4} \,\rm K$) in red hues and cool gas ($T \leq 10^{4} \,\rm K$) in blue hues. The black dashed circle marks 20 kpc. We see that \small{Jet AGN} and \small{Spawn AGN} models significantly affect the hot gas structure and diminish the cool gas extent, while \small{Push AGN} predominantly disrupts cold gas in the inner region, explaining the thin disk substructure differences between both AGN models (see Figure \ref{fig:2dhist_m12i} or \ref{fig:2dhist_m12f}). 
\label{fig:gas_temp_massplot}}
\end{figure*}

\subsection{Comparing Feedback Implementations \label{subsec:jetvspush}}

As discussed in the Introduction, the three AGN feedback implementations we use here have the same energetics and efficiencies for the main feedback channels (cosmic rays, radiation, and winds).  The only difference is that they inject the mass, energy, and momentum differently. As such, the differences we witness in final morphologies, as well as the differences in star formation suppression, relate to how the energy is imparted and the resulting effective efficiencies of each feedback model. As a reminder, the \small{{Push AGN} implementation deposits mass, energy, and momentum directly into existing gas cells in an isotropic fashion. Both \small{Jet AGN} and \small{Spawn AGN} implementations instead deposit these quantities through the creation of new, high-resolution energetic particles. \small{Spawn AGN} injects these particles isotropically, while \small{Jet AGN} deposits them in the angular momentum direction of the accreted material.  

Of the three models, \small{Jet AGN} has the most dramatic effect on the visual morphology (Figures \ref{fig:mockm12i} and \ref{fig:mockm12f}), the most star formation suppression (Figures \ref{fig:SMHMplot} and \ref{fig:circvslbt}), and produces the least massive thin-disk components (see Figure \ref{fig:circplot}) and the smallest radii disks (see Figures \ref{fig:2dhist_m12i} and \ref{fig:2dhist_m12f}). The \small{Push AGN} implementation has the least dramatic effect on visual morphology and star formation suppression in both runs. The two \small{Push AGN} runs even produce star-forming thin-disk components at $z=0$, but these components are much lower in total mass and in surface density than in the \small{No AGN} cases. The one \small{Spawn AGN} run provides similar results to \small{Jet AGN} run. Its non-star-forming disk is more extended than its counterpart in \small{Jet AGN}, but it is not as extended as \small{Push AGN} (see Figure \ref{fig:2dhist_m12i}).    

Figure \ref{fig:gas_temp_massplot} provides some insight into these differences by presenting mass surface density plots for both cool/cold gas ($T< 10^4 \,\rm K$, blue) and warm/hot gas ($T > 10^4 \,\rm K$, red) at various steps in lookback time for each of the runs. The $z=0$ images are on the far left and the earliest images on the right, with lookback time labeled along the top rows.  Each box spans 150 kpc on a side, and thus shows both the inner ISM (mostly blue/cool) and outer CGM (mostly red/warm) of the system. Each AGN model occupies a separate row, as labeled. The m12i cases are shown above the m12f runs. 

It is clear from Figure \ref{fig:gas_temp_massplot} that the \small{Push AGN} implementation is affecting its outer, hot CGM the least compared to the other AGN runs. The hot gas structure in the \small{Push AGN} cases is comparable to the runs without AGN. As shown in Appendix \ref{app:ISM_CGM}, the hot CGM of m12f is actually {\em more massive} in Push AGN than in its non-AGN. Though this may suggest some degree of preventative feedback keeping the hot halo from cooling, much of this occurs prior to the time of inner CGM virialization \citep{Myrtaj2026}. It is more likely this difference is associated with the fact that the star formation rate of this system was significantly reduced compared to the AGN run, especially just after spin-up.  Reducing the star formation rate at this time likely results in fewer {\em stellar} feedback-driven outflows into the IGM, allowing the CGM to maintain more mass. 

We do see significant differences in the {\em inner} cool gas structure of the \small{Push AGN} runs relative to the non-AGN cases. This is perhaps not surprising, given that this approach injects its feedback more locally around the black hole, which allows for strong effects centrally but may be less effective at impacting the outer hot gas. 

Images for both the \small{Jet AGN} and \small{Spawn AGN} runs, on the other hand, show a clear depletion of warm and cool gas relative to the runs without AGN. Those differences become visually apparent after the spin-up times (and associated SFR suppression times) for each of these objects: after $\sim 8$ Gyr for m12i and after $\sim 10$ Gyr lookback for m12f. Spawning particles clearly results in more efficient feedback in these simulations. For the single comparison case we have, the \small{Jet AGN} (collimated injection) implementation appears to create a larger effect on the gas than \small{Spawn AGN}. It is possible that by collimating the injection (pointed out of the disk plane), the energy and momentum can reach farther into the CGM, perhaps leading to more efficient overall gas loss. \small{Spawn AGN} imparts the feedback isotropically,  which may allow the energetic particles to interact more directly with gas as it builds up in the disk plane, allowing less energy and momentum to reach gas at large radii.

\subsection{Spin Up and Effective AGN Feedback \label{subsec:spinup_and_AGN}}

In all of our runs, the galaxy undergoes a transition where gas and new stars within 20 kpc initially have circularities that are fairly radial, and then begin to rise steadily towards more circular, disk-like orbits (Figure \ref{fig:circvslbt}).  We have noted above that star formation suppression in all AGN runs begins to set in after this ``spin-up" transition. Our observed relationship between the spin-up time and the onset of efficient negative AGN feedback suggests that either 1) the two events share a common cause, or 2) coherent rotation in the gas enables more effective AGN feedback. 

Although the cause of spin-up is unclear, \citet{Myrtaj2026} found, using runs without AGN, that the spin-up time tends to occur just after the settling of the galactic center of mass motion. It could be, then, that having a stable center creates conditions where energy and momentum injection from the SMBH acts more effectively on the surrounding gas.

Regardless of {\em why} the galaxy spins up, it may be that the change in kinematic structure of the gas near the galactic center is what allows AGN feedback to be more effective. For example, when the gas is rotationally supported, it will tend to accrete along a direction perpendicular to its angular momentum.  This may provide a channel (parallel to the angular momentum direction) for energy and momentum to escape and act on gas reservoirs at larger radii.

Another interesting result related to spin up is that in all the AGN runs, we find that galaxies spin up sooner than the non-AGN counterparts (see Table \ref{tab:one} and Section \ref{subsec:morph_overtime}. This observation is based on only five runs, so it could be a statistical fluctuation.  Alternatively, the presence of the massive central black hole could potentially expedite the settling of the center of mass motion, which could then enable spin-up earlier in these cases. More simulations of this kind will be needed to test this possibility.

\subsection{Lenticular Formation} 
\label{subsec:lenticular}} 

Notably, our work demonstrates a formation pathway for lenticular galaxies under secular AGN feedback.  Specifically, both \small{Jet AGN} and \small{Spawn AGN} runs for m12i produce galaxies with featureless, non-star-forming disk components at $z=0$ (see the bottom row of Figure \ref{fig:mockm12i}). The origin of these lenticular morphologies has little to do with mergers.  In both cases, AGN feedback begins to suppress star formation after the galaxy ``spins up" and starts producing a disk, and then completely evacuates the gas at late times, after which star formation stops. Some stars form on disk orbits before the time of complete quenching.  Ultimately, however, the disk stars stop forming in these two AGN runs as the whole galaxy begins to quench at a lookback time of $\sim 4$ Gyr.   The pre-existing disk stars simply age and become red, while kinematic heating washes away residual spiral features.  

We have thus presented a formation pathway for lenticular morphologies that is driven by gradual, ongoing AGN-driven feedback, which provides complete quenching only after some early disk formation has occurred.  This pathway is qualitatively consistent with early ideas from \citet{Larson1980DiskGalaxiesS0} on the origin of S0 galaxies.  For an informative discussion of lenticulars as faded spirals, see \citet{KormendyBender2012ParallelSequence}.

\subsection{Mergers \label{subsec:mergers}}
One of our initial conditions, m12f, has significantly more late-time minor mergers than m12i, and therefore provides an interesting test case for understanding how mergers affect morphologies differentially in runs with and without AGN. Specifically, m12f has two significant (LMC-like) mergers that occur roughly five to six billion and one billion years before the present day  \citep[both with halo masses of $\sim 10^{11}$ M$_\odot$, see][]{zhang2026night}.   

In non-AGN cases, these mergers leave behind signatures in the phase-space structure of stars and at faint surface brightness \citep{Donlon2024MNRAS}, but they do {\em not} induce any significant features in the circularity evolution of young stars or cold gas in the main progenitor (blue lines, bottom left panels of Figure \ref{fig:circvslbt}).  Moreover, they do not prevent the formation of a thin disk at $z=0$ with flocculent spiral structure (see the left panel of Figure \ref{fig:mockm12f}). 

The same mergers have a more dramatic effect on the evolution and late-time morphology of the main progenitor of m12f in the two runs with AGN feedback. For example, the young star circularity evolution shows two strong dips at times corresponding to these mergers (green and orange lines, bottom left panels of Figure \ref{fig:circvslbt}).~\footnote{Note that the precise timing of these mergers is not the same from run to run.  This is not unexpected given the dynamical impact of AGN feedback on mass distributions in and around the galaxies.} The most recent merger event actually triggers a late-time starburst in the \small{Jet AGN} run of m12f (see right bottom panel of Figure \ref{fig:circvslbt}). We saw similarly timed features in the total spheroidal and thick disk mass evolution in Figure \ref{fig:morph_massplot}. 

Finally, the visual morphologies of the two m12f AGN runs clearly betray recent merger activity (middle and right panels of Figure \ref{fig:mockm12f}). The \small{Push AGN} case produces a clearly disturbed disk, while the \small{Jet AGN} run is even more disordered in overall shape and displays a bright, young, disk-like component at small radii, which appears to have formed from the late-time, gas-rich merger. 

The explanation for these differences lies primarily in the differential effects that AGN feedback has on the gas and stellar content of the main progenitor; the merging galaxies are so small that the AGN does not strongly affect their baryonic content.  In the non-AGN case, the associated mergers are quite minor, with baryon mass ratios well below $\sim 1/10$, owing to the relatively large denominator mass of the main progenitor.  In the AGN runs of m12f, however, the main galaxy has a much lower stellar mass ($\sim 4$ times lower at late times) and gas mass ($\sim 2$ times lower for much of its history). This means that the mergers are closer to $\sim 1/10$ in gas and stellar mass, resulting in more significant dynamical responses.  

\section{Summary} \label{sec:summary}
We have analyzed seven cosmological zoom-in FIRE-2 simulations of Milky Way mass galaxies with and without AGN feedback to probe the relationship between star formation suppression and galaxy morphology. We specifically studied two initial conditions, m12i and m12f. The first has a fairly quiescent merger history, and the second has more mergers, with two LMC-size mergers in the last six billion years. Each of them is run without AGN feedback and again with up to three different AGN feedback implementations.  These implementations have the same energetics and feedback efficiencies for mechanical winds, cosmic rays, and radiation pressure (see Section \ref{subsec:agnparams}). \small{Push AGN} imparts feedback energy directly into gas cells in an isotropic sphere, \small{Jet AGN} imparts the energy as a stream of energetic particles along a collimated path, and \small{Spawn AGN} imparts the energy as a stream of energetic particles in an isotropic sphere (see Section \ref{subsec:sims}).  

Both runs without AGN feedback produce prominent, thin, star-forming spiral disks at z = 0. We find that all AGN feedback models reduce the gas mass in the central region of the galaxy, produce less overall star formation, and make galaxies with lower thin-disk stellar mass fractions and higher spheroidal mass fractions (See Table \ref{tab:one} and Figures \ref{fig:mockm12i}, \ref{fig:mockm12f},  \ref{fig:SMHMplot}). In two cases with AGN feedback (m12i with \small{Jet} and \small{Spawn AGN}), we produce quenched, lenticular galaxies. 

To explore the origin of the morphological differences in the AGN runs, we quantify the morphological properties of stars as they form and evolve using particle circularity, $\epsilon = j_{\rm z}/j_{\rm c}(\rm E)$, and classified components as spheroidal ($\epsilon \leq 0.2$), thick disk ($0.8 \geq \epsilon > 0.2$), and thin disk ($1.0 \geq \epsilon > 0.8$), as shown in Figures \ref{fig:circplot}, \ref{fig:2dhist_m12i}, and \ref{fig:2dhist_m12f}. 

By tracking the median circularities of newly-formed star particles over time, we show that each of our runs undergoes three phases of kinematic evolution: an early phase with radial, spheroidal-type orbits, which subsequently ``spins up" into a second, thick-disk phase with coherently rotating material, and finally, a late-time ``thin-disk" phase where stars and gas have mostly circular orbits (see Figure \ref{fig:circvslbt}).  This result has been shown elsewhere for runs without AGN feedback \citep{Stern2021, Yu2021, Hafen2022, Yu2023, Gurvich2023, Myrtaj2026}, but here we show that these same transitions occur in runs with AGN feedback. Interestingly, we find that, consistently, the runs with AGN spin up slightly ($\sim 0.8$ Gyr) earlier than the runs without AGN feedback. 

The key finding in our paper is that, in all cases, star formation in the AGN runs is suppressed relative to the non-AGN runs  {\em after} the spin-up time (see the right column of Figure \ref{fig:circvslbt}). This means that disk formation is preferentially suppressed by AGN feedback, while the early spheroid-forming phase is very similar to runs without AGN. Importantly, in most cases, the star formation that does occur at late times in the runs with AGN {\em does} occur on thin-disk-like orbits.  The primary reason that the disk fractions are lower in these runs is that the star formation rates during this phase of formation are much lower.

Interestingly, around the time of ``spin up", BH accretion also peaks, undergoing steady gas accretion across all AGN models. The effects of AGN feedback `turn on' only after this period, marking a time of star formation suppression in the central region of the galaxy under AGN feedback. The same star formation suppression coincides with a reduction in cool ISM gas compared to the non-AGN cases (Figure \ref{fig:massplot}). 
Beforehand, the galaxy evolves similarly regardless of the AGN model.  

 We emphasize that spin-up marks a transition for AGN feedback in the star formation rate evolution (Figure \ref{fig:circvslbt}), the cool ISM gas evolution (Figure \ref{fig:massplot}), and CGM gas evolution (Figure \ref{app:ISM_CGM}).  Since AGN feedback begins to take effect only after the system becomes angular momentum-supported, it continuously imparts energy and momentum to the gas reservoir throughout disk formation. This results in a higher spheroidal fraction at $z=0$. 

Additionally, the choice of energy injection and geometry affects the disk substructure and the efficiency of the SMBH feedback and accretion. The \small{Jet} implementation is most efficient at suppressing star formation.  It also produces the smallest thin-disk components and highest spheroidal mass fractions in all runs. In contrast, the \small{Push} implementation appears less efficient at feedback, resulting in high gas mass in the central region of the galaxy and higher accretion owing to its higher $M_{\rm BH}$ and $\dot{M}_{\rm BH}$ and total $M_{\rm gas}$ compared to \small{Jet}. In the middle, the \small{Spawn AGN} implementation is more efficient than \small{Push AGN} in suppressing star formation, but morphologically produces a slightly lower spheroidal fraction. In both star formation suppression and thin-disk suppression, it is less effective than Jet, indicating that isotropic injection is the less efficient geometry. 

As mentioned above, in two instances -- both in the quiescent m12i initial condition -- we have AGN runs that produce quenched, lenticular galaxies with no cool gas at $z = 0$. Specifically, the \small{Jet AGN} and \small{Spawn AGN} runs of m12i produce galaxies with featureless, non-star-forming disk components at $z=0$ (see the bottom row of Figure \ref{fig:mockm12i}).  In these two cases, AGN feedback begins to suppress star formation after the galaxy spins up and starts producing a disk, and then completely evacuates the cool ISM gas at late times. Star formation shuts down almost completely in this galaxy about 4 billion years ago. The pre-existing disk stars simply age and become red, while kinematic heating washes away residual spiral features.  This pathway to lenticular formation is consistent with the ``faded spirals" picture of S0 galaxy formation \citep{Larson1980DiskGalaxiesS0}. In contrast, the \small{Push AGN} run of m12i produces a galaxy with very low star formation, confined to a blue outer ring (see Figure \ref{fig:mockm12i}), which may be on its way to quenching. Interestingly, \citet{Smith2022} observe that ring structures are more common in green-valley (transition) galaxies compared with their red and blue counterparts.  

The other differences we note with the AGN runs are the final, ``thin disk" kinematic phase of star and gas orbits.  In runs without AGN, it is well established that the circularity distributions of young stars and cool gas in Milky Way-like FIRE-2 galaxies are highly peaked towards $\epsilon = 1$ at late times \citep[][and references therein]{Myrtaj2026}.  We see this in the blue lines on the left panel in Figure \ref{fig:circvslbt}, which show that the {\em median} value is $\epsilon \simeq 1$ at late times.  \citet{Stern2021} and \citet{Hafen2022} explain this phenomenon by showing that after the virialization of the inner CGM, infalling gas has time to mix and fall into the galaxy with extremely well-aligned angular momentum vectors before joining the disk. In contrast, although the circularities of young stars in the AGN runs (Figure \ref{fig:circvslbt}) tend to have thin-disk-like orbits at late times ($\epsilon \gtrsim 0.8$), they do {\em not} remain peaked at almost perfectly circular orbits $\epsilon \sim 1$.  This difference is even more apparent in the cold gas kinematics (middle panels of Figure \ref{fig:circvslbt}) where the orbits betray a distribution that is even less ``thin."  In m12i, in particular, which is not affected by mergers, these differences could be driven by the AGN feedback effect on the hot CGM structure, which may prevent coherent mixing and alignment of angular momentum in the infalling material prior to joining the disk. More work will be required to test this effect more directly by tracking particle orbits as in \citet{Hafen2022}.

The results of this paper demonstrate a mechanistic case study of how AGN feedback injection impacts morphology. We focused on providing general insights into the physics driving morphological changes under AGN feedback. We note that the broad-morphology features in our AGN samples are robust over cosmic time, with slight deviations attributable to minor mergers.

An important caveat to the results from this work is the small sample size, which may not be representative of the parameter space of detailed morphologies associated with AGN feedback.  Further investigation into the physical motivations and timings of these events is required.

Missing from our analysis are the effects of AGN feedback under a major merger and the role AGN feedback has on the dynamics of a merger. In a follow-up paper, we examine m12q, another Milky Way mass galaxy that undergoes a major gas-rich merger at late times, which produces a disk under our AGN feedback models than without AGN. 

\begin{acknowledgments}
PF is supported by a GAANN Fellowship with funding from the US Department of Education under contract P200A240014 and NSF grant AST-2408246. JSB is supported by NSF grant AST-2408246. SW received support from the NASA RIA grant 80NSSC24K0838. JM is funded by a Pomona College Large Research Grant. FJM is funded by the National Science Foundation (NSF) Math and Physical Sciences (MPS) Award AST-2316748, and AW received support via NSF CAREER award AST-2045928.
\end{acknowledgments}

\bibliography{citations}{}
\bibliographystyle{aasjournalv7}

\appendix

\begin{figure}[!htb]
\centering
\includegraphics[width=0.48\textwidth]{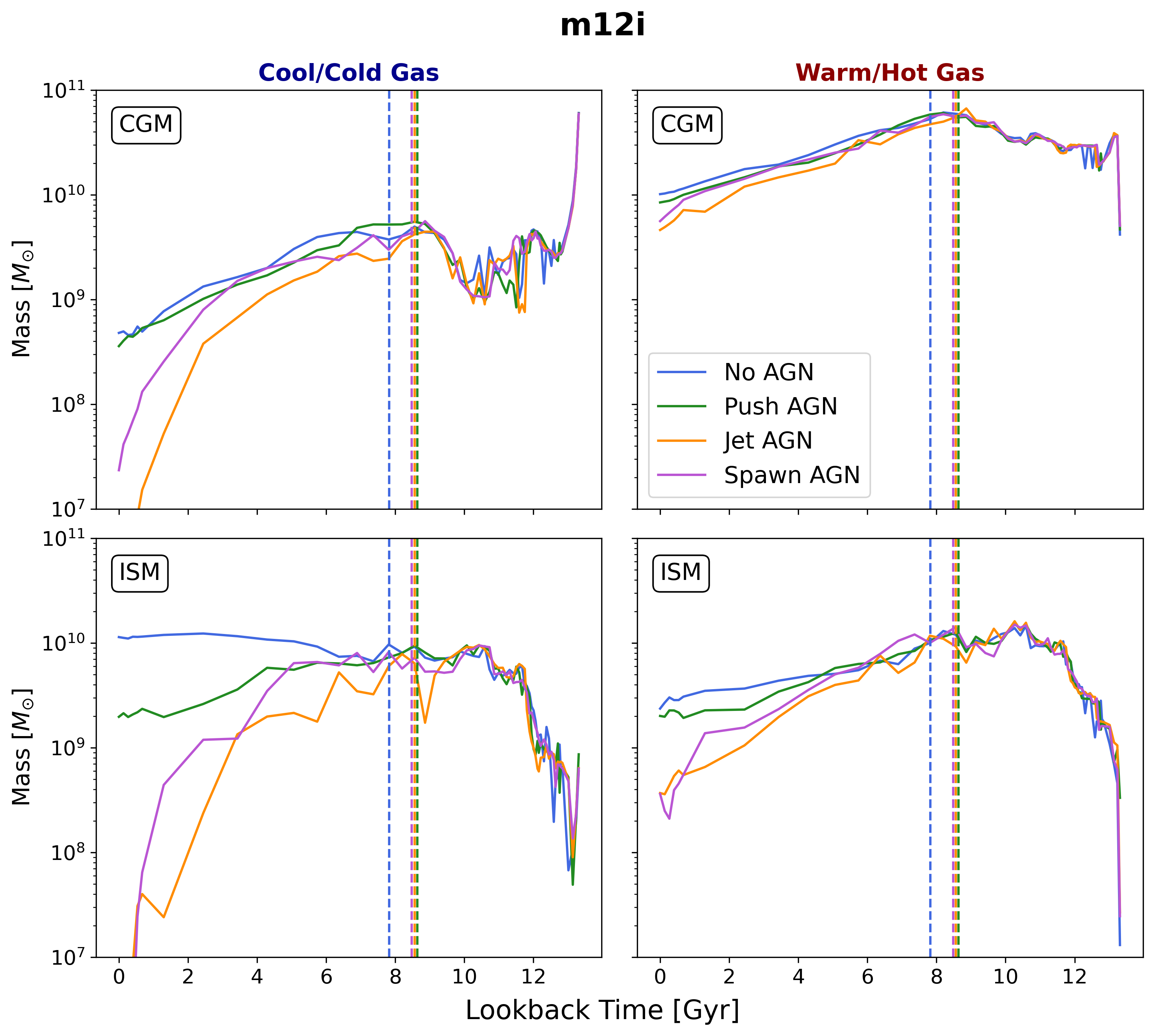}
\caption{CGM and ISM evolution in m12i.
Shown is the cool/cold ($\rm T<10^4 \,\rm K$, left column) and warm/hot ($\rm T>10^4 \,\rm K$, right column) gas mass within the CGM ($20 \,\rm kpc < r < 100\, \rm kpc$, top row) and ISM ($r < 20$ kpc, bottom) for each m12i run as a function of lookback time. Spin-up times for each run are shown by the vertical dotted lines, matched by the color code of the runs in the legend. We see that the primary differences between the AGN runs and non-AGN runs occur in the cool ISM after spin-up. The hot CGM also shows some differences from run to run, but not as dramatically as the cool ISM. 
\label{fig:m12i_gas}}
\end{figure}

\begin{figure}[!htb]
\centering
\includegraphics[width=0.48\textwidth]{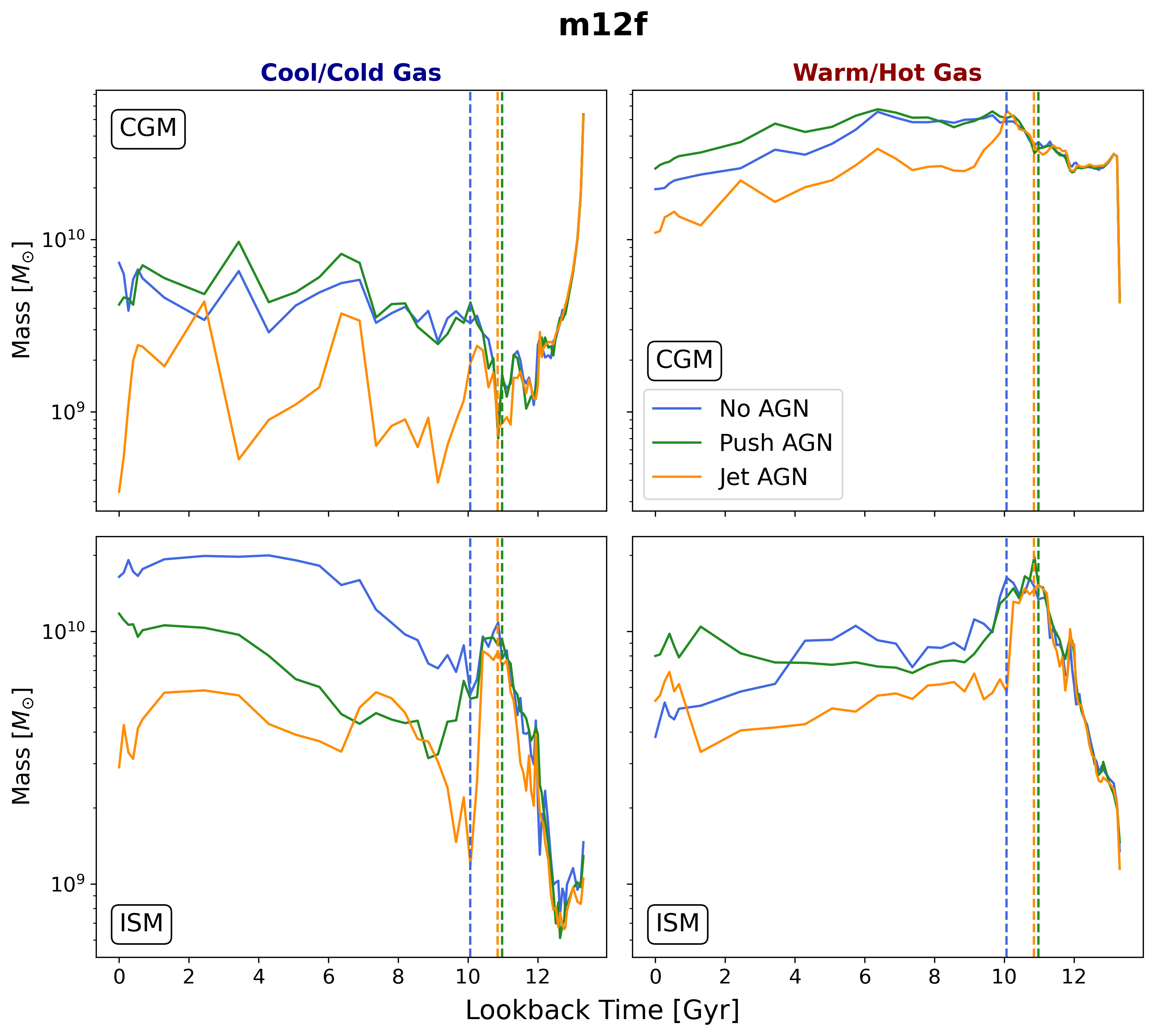}
\caption{CGM and ISM evolution in m12f.  The panels mirror those in Figure \ref{fig:m12i_gas}. As we saw with m12i, the primary differences are seen after spin-up in the cool ISM.  
\label{fig:m12f_gas}}
\end{figure}

\section{Evolution of the ISM \& CGM}
\label{app:ISM_CGM}

 Here we show how cool and warm gas in the ISM and CGM of each of the runs evolves with lookback time and compare results with and without different AGN models. The top row of Figure \ref{fig:m12i_gas} shows the mass in cool/cold ($\rm T<10^4 \,\rm K$, left column) and warm/hot ($T \rm > 10^4 \,\rm K$, right column) gas within the CGM ($20 \,\rm kpc < r < 100\, \rm kpc$, top row) as a function of lookback time for m12i. The bottom row shows the same for the ISM ($\rm r < 20$ kpc, bottom) as a function of time for m12i. Figure \ref{fig:m12f_gas} shows the same panels for m12f.

For the specific case of m12i, which has a fairly quiescent merger history, under AGN feedback, the gas mass decreases continuously as the SMBH mass ($M_{\rm BH}$) increases. Without AGN feedback, the gas mass within 20 kpc stays relatively steady, reflecting an equilibrium between star formation and fresh gas accretion.  The CGM also shows a gradual reduction in mass, with only mild differences between the AGN runs and non-AGN runs. 

The m12f case is similar, though here we see more significant differences in the warm CGM evolution. Interestingly, the hot halo in the Push AGN run is actually more massive in this case than in the non-AGN run, suggesting some degree of preventative cooling. However, it seems more likely that this has to do with the fact that the star formation rate of this system was significantly reduced compared to the non-AGN run, especially just after spin-up. This likely results in less {\em stellar} feedback-driven outflows into the IGM.

The Jet AGN run has less hot CGM than either of the other runs, suggesting that more of the gas has escaped to the IGM in this run.

However, the overall stellar masses of the AGN runs are significantly reduced in both halos compared to the non-AGN runs. In the case of m12f, the stellar mass drops from $8.3 \times 10^{10}\,\rm M_\odot$ without AGN to $\sim 2-3 \times 10^{10}\,\rm M_\odot$ with AGN. For m12i, the drop is similar, from $5.9 \times 10^{10}\,\rm M_\odot$ to $\sim 2.5 - 3.2 \times 10^{10} \,\rm M_\odot$. One clear conclusion is that a significant mass in baryons has been lost to the IGM as a result of AGN feedback in all cases. That is, the bulk of the feedback is not ``maintenance heating" of the CGM.

\end{document}